\documentclass[showkeys,showpacs,superscriptaddress,nofootinbib,aps,prd]{revtex4-2}
\usepackage{amsmath}
\usepackage{amssymb}
\usepackage{graphicx}
\usepackage{xcolor}
\usepackage{float}
\usepackage{physics}

\allowdisplaybreaks

\begin{document}

\title{
Nonperturbative Dyson--Schwinger equations in QCD: a first approximation 
}
\author{
Vladimir Dzhunushaliev
}
\email{v.dzhunushaliev@gmail.com}
\affiliation{
Department of Theoretical and Nuclear Physics,  Farabi University, Almaty 050040, Kazakhstan
}
\affiliation{
Institute for Experimental and Theoretical Physics, Farabi University, Almaty 050040, Kazakhstan
}

\affiliation{Academician J.~Jeenbaev Institute of Physics of the NAS of the Kyrgyz Republic, 265 a, Chui Street, Bishkek 720071, Kyrgyzstan}

\author{Vladimir Folomeev}
\email{vfolomeev@mail.ru}

\affiliation{
Institute for Experimental and Theoretical Physics, Farabi University, Almaty 050040, Kazakhstan
}

\affiliation{Academician J.~Jeenbaev Institute of Physics of the NAS of the Kyrgyz Republic, 265 a, Chui Street, Bishkek 720071, Kyrgyzstan}

\begin{abstract}
We consider a procedure for nonperturbative quantization based on an infinite system of nonperturbative Dyson--Schwinger equations. We propose a first approximation for truncating this infinite system of equations in the static case. The main ideas underlying this approximation are as follows: (a) the two-point Green's functions of vacuum gauge fields are factorized by introducing scalar fields; (b) in the non-vacuum case, the degrees of freedom of the $SU(3)$ gauge field can be divided into ``almost classical'' and ``almost quantum'' degrees of freedom; (c) the four-point Green's functions of the gauge fields are represented as bilinear combinations of two-point Green's functions; (d) an approximation for the three-point Green's function describing the interaction between quarks and gauge fields is proposed, which leads to the splitting of the original Dirac equation for the quark-field operators into two equations, one describing the expectation value of the fermion field and the other acquiring a nonlinear term and serving to describe a condensate composed of sea quarks bound by the vacuum gauge field.
We consider the choice of gauge appropriate for this approximation to nonperturbative quantization. We point out that the emergence of a nonlinear Dirac equation within this approximation may lead to a mass gap in the energy spectrum of solutions of the corresponding systems of equations. The issue of the emergence of ``dimensional transmutation'' within this approximation is also discussed. We provide arguments in favor of the statement that the ``closure constants'' appearing in the finite truncation and giving rise to ``dimensional transmutation'' should survive in the transition to the infinite system of Dyson--Schwinger equations. An analogy between nonperturbative quantization and the stochastic theory of turbulence is also discussed.
\end{abstract}

\pacs{}

\keywords{nonperturbative quantization, Dayson--Schwinger equations, quantum chromodynamics, nonlinear Dirac equation, closure constants}
\date{\today}

\maketitle 

\section{Introduction}
\label{intro}

Quantum field theory based on perturbative calculations using Feynman diagrams has achieved remarkable success in quantum electrodynamics (QED) and the electroweak theory. Its predictions are in excellent agreement with experimental data to a very high degree of precision. Nevertheless, the founders of quantum field theory were cautious in assessing the foundations of perturbative quantum field theory and apparently believed that further work was needed to clarify the issues arising within the perturbative framework.

R.~Feynman stated the following in this regard:
``What the three Nobel Prize winners did ... was to get rid of the infinities in the calculations. The infinities are still there, but now they can be skirted around ... We have designed a method for sweeping them under the rug'' \cite{Feynman}. 

Paul Dirac himself also wrote with similar tune \cite{Dirac}:
“Hence most physicists are very satisfied with the situation. They say: ``Quantum electrodynamics is a
good theory, and we do not have to worry about it any more.'' I must say that I am very dissatisfied with
the situation, because this so-called ``good theory'' does involve neglecting infinities which appear in its
equations, neglecting them in an arbitrary way. This is just not sensible mathematics. Sensible mathematics
involves neglecting a quantity when it turns out to be small—not neglecting it just because it is infinitely great
and you do not want it!''. 

L.~Landau  et al.~\cite{Landau:1} wrote on this subject: 
``Although at present there exist methods to remove these singularities (regularization), which clearly lead to correct results, such method of action has the artificial nature. The singularities arise in the theory due to the pointlike interaction described by  delta functions (operators of interacting fields are taken at one point).'' In Refs.~\cite{Landau:2,Landau:3,Landau:4}, he and his co-authors study this question trying to remove such singularities in QED.

Another argument in favor of this viewpoint is the position of W. Heisenberg, which he articulated in Ref.~\cite{heis} (chapter 2, \S2): ``The initial or the final states are, by definition, states in which the interaction between the particles can be neglected. Such states, and only such states, can be constructed by applying sums of products of asymptotic operators on the vacuum. They are usually supposed to form a complete set of states in the sense that any process of interaction will finally, when the interaction has become negligible, lead to a state within this set. Even if this is true, the state of interaction itself may belong to a wider group of states which cannot be represented by the asymptotic states alone.'' 
This viewpoint of W. Heisenberg should be understood to mean that the properties of strongly interacting field operators fundamentally differ from those of weakly interacting fields; therefore, the description of strongly interacting fields must be inherently distinct from the methods developed for weakly interacting fields.

All these statements by R.~Feynman, P.~Dirac, L.~Landau, and W.~Heisenberg suggest that perturbative approaches based on a small coupling constant are not yet a fully understood mathematical construction. We have simply been fortunate that this method works remarkably well in the range accessible to our experimental capabilities. However, one cannot be certain that it will remain valid at all energy scales of particle interactions, at arbitrary distances, in curved spacetime, and so on.
For example, quantum chromodynamics (QCD) presents such fundamental problems as confinement, the mass gap, and many others, which apparently cannot be resolved on the basis of perturbative calculations. In Ref.~\cite{Dzhunushaliev:2026rrt}, arguments are presented suggesting that, in the vicinity of a black-hole event horizon, the canonical anticommutation relations for fermions may be modified, leading to the emergence of a nonperturbative fermionic vacuum.

All the statements and arguments presented above support the view that, in order to achieve a complete understanding of quantum field theory, it is necessary to formulate a nonperturbative quantum field theory. The need for such a theory is particularly evident in QCD, where numerous problems arise from the inability to employ perturbative methods that work remarkably well within the Standard Model.

At present, there exists a well-established and effective method of nonperturbative quantization, namely, lattice methods based on numerical evaluations of the path integral in Euclidean space. The main drawback of this approach is that all calculations are numerical, which means that the physical processes underlying the resulting phenomena are not always transparent.
It is important to emphasize that by nonperturbative quantization we mean a method that, in principle, allows one to calculate any quantum-field-theoretic quantity of interest, such as Green's functions, quantum expectation values, quark and gluon condensates, and so on. This situation can be compared with classical field theory, where field equations (the Maxwell, Yang--Mills, Einstein equations, and so on) provide, in principle, a means of obtaining solutions describing a given physical system. In perturbative quantum field theory, the corresponding framework is provided by Feynman diagram techniques together with the renormalization procedure.
For the nonperturbative quantization of a nonlinear spinor field, W.~Heisenberg \cite{heis} proposed using an operator equation, or, equivalently, an infinite system of nonperturbative Dyson--Schwinger equations for all Green's functions. We assume that such an approach can be applied to any nonlinear field theory and employ it in QCD by truncating the infinite system of nonperturbative Dyson--Schwinger equations to a finite system involving the lowest-order Green's functions.
Let us emphasize the distinction between this nonperturbative approach and the corresponding treatment in perturbative quantum field theory. In the latter case, the perturbative Dyson--Schwinger equations are solved by expanding in the small coupling constant and employing Feynman diagrams.

Nonperturbative quantization based on an infinite system of nonperturbative Dyson--Schwinger equations was apparently first proposed by W.~Heisenberg for the nonlinear Dirac equation; see the textbook \cite{heis}. In Ref.~\cite{Bender:1999ek}, a truncation scheme for the Schwinger--Dyson equations was employed to perform calculations in $g\phi^4$ scalar quantum field theory. In Ref.~\cite{Frasca:2019ysi}, the Dyson--Schwinger equations for QCD describing the one- and two-point Green's functions were derived, and it was shown that, in the ’t Hooft limit, they can be reduced to a form suitable for analysis. In Ref.~\cite{DiGiacomo:2000irz}, a formalism based on gauge-invariant nonlocal correlators in non-Abelian gauge theories was considered. It was shown that many nonperturbative aspects of gluodynamics and QCD of phenomenological interest can be described in terms of correlators of non-Abelian field-strength tensors.

Our approach to the approximate solution of the nonperturbative Dyson--Schwinger equations involves a number of issues that require further investigation:
\begin{itemize}
\item When truncating an infinite system of equations to a finite one, an appropriate truncation scheme must be chosen. One may either neglect the higher-order Green's functions, express them in terms of lower-order Green's functions, or employ some other procedure for eliminating the higher-order Green's functions. This issue is well known in turbulence modeling as the closure problem; see, e.g., the textbook \cite{Wilcox}. 
\item When truncating the nonperturbative Dyson--Schwinger system, various types of Green's functions arise, for example,
$\left\langle \hat{A}^B_\mu \hat{A}^C_\nu \right\rangle$,
$\left\langle \hat{\bar{\psi}} \hat{A}^B_\mu \hat{\psi} \right\rangle$,
and so on. Our main assumption in this case is the following: \textit{the approximate description of integer-spin bosonic fields (in our case, gauge fields) is achieved by introducing suitable bosonic fields (a scalar field $\phi$ and/or a vector field $A^a_\mu \in G \subset SU(3)$), while the approximate description of fermions is achieved by means of a spinor field.}
\item When truncating the Dyson--Schwinger equations, the question of the gauge invariance of the resulting equations arises and requires careful investigation. 
\item A major difficulty is the choice of an appropriate approximation for Green's functions that involve both gauge-field and fermionic operators, such as the three-point Green's function
$\left\langle \hat{\bar{\psi}} \hat{A}^B_\mu \hat{\psi} \right\rangle$.
For example, when $\left\langle \hat{A}^B_\mu \right\rangle = 0$, one cannot assume that
$
\left\langle \hat{\bar{\psi}} \hat{A}^B_\mu \hat{\psi} \right\rangle
\approx
\left\langle \hat{\bar{\psi}} \hat{\psi} \right\rangle
\left\langle \hat{A}^B_\mu \right\rangle,
$
since this approximation would imply that the three-point Green's function vanishes,
$
\left\langle \hat{\bar{\psi}} \hat{A}^B_\mu \hat{\psi} \right\rangle \approx 0.
$
However, this cannot generally be the case, since the gauge field $A^B_\mu$ and the fermionic field $\psi$ are coupled through the Yang--Mills and Dirac equations. This coupling induces correlations between the fields and, consequently, their quantum expectation value need not vanish.
\item The nonperturbative Dyson--Schwinger system constitutes an infinite system of partial differential equations. The properties of such a system may \textit{differ fundamentally from those of a finite system of equations}. For example, solutions of an infinite system of equations may contain dimensional constants that are absent from the original system of equations, a phenomenon known as \textit{dimensional transmutation}. In QCD, for instance, there is the scale $\Lambda_{\text{QCD}}$, which may be conjectured to emerge precisely upon integrating the infinite system of equations.
This raises the following question: how can such constants arise in a finite truncated system of Dyson--Schwinger equations when the original infinite system is truncated to a finite one? A possible answer is the following. Upon truncation, higher-order Green's functions are approximated by polynomial combinations of lower-order Green's functions. In such an approximation, certain coefficients necessarily appear in the polynomial. For example, a four-point Green's function $G_4$ can be approximated in terms of the two-point Green's function $G_2$ as
$
G_4 \approx G_2^2 + \lambda_1 G_2 + \lambda_2,
$
where the dimensional constants $\lambda_{1,2}$ are referred to as ``closure constants.'' If we assume that such coefficients survive in the limiting transition to the infinite system of equations, this would imply that dimensional constants absent from the original fundamental field equation emerge in the infinite system of Dyson--Schwinger equations.
\end{itemize}

The paper is organized as follows. In the Introduction, Sec.~\ref{intro}, we discuss some remarks by five Nobel laureates concerning problems associated with the procedure of perturbative quantization and outline several issues arising in nonperturbative quantization based on the nonperturbative system of Dyson--Schwinger equations for all Green's functions. In Sec.~\ref{YMD_operator_eqn}, we present the idea of nonperturbative quantization proposed by Heisenberg. In Sec.~\ref{vacuum_one_field}, we consider the vacuum case. In Sec.~\ref{two_fields}, we discuss an approximation involving ``almost classical'' and ``almost vacuum'' degrees of freedom. In Sec.~\ref{quark_approx}, we introduce the approximations for fermions (quarks) required to obtain the averaged Dirac equation and the currents appearing on the right-hand sides of the Yang--Mills equations. In Sec.~\ref{final_eqns}, we summarize all the equations describing various physical situations and discuss their physical properties. In Sec.~\ref{gauge_inv}, we specify the gauge choice adopted in our approximation. In Sec.~\ref{turb_QCD}, we discuss a possible deep connection between nonperturbative quantization and the stochastic theory of turbulence. Finally, in Sec.~\ref{conclusions}, we summarize the proposed procedure for truncating the infinite system of nonperturbative Dyson--Schwinger equations and discuss the remaining issues. All the lengthy calculations required in Secs.~\ref{vacuum_one_field}--\ref{quark_approx} are collected in Appendices \ref{one_scalar_field}--\ref{Dirac_eqn}.

\section{Yang--Mills--Dirac operator equations and nonperturbative Dayson--Schwinger equations in QCD 
}
\label{YMD_operator_eqn}

According to W.~Heisenberg~\cite{heis}, the procedure of nonperturbative quantization consists in writing out quantum averages of the Yang--Mills and Dirac operator equations. Consider the operator Yang--Mills--Dirac field equations, 
\begin{align}
	\hat F^{A \mu \nu}_{; \nu}	= & - \hat j^{A \mu} = 
	- \hat{\bar{\psi}} \lambda^A \gamma^\mu \hat{\psi},
\label{oprtr_YM}\\
	D_\mu \hat{\psi} -m \hat{\psi} = & 0 , 
\label{oprtr_D}
\end{align}
where $A, B,C,D=1,2, \ldots 8$ are color indices and $\mu, \nu = 0, 1, 2, 3$ are spacetime indices; $D_\mu \psi = \partial_\mu \psi  - \imath \frac{g}{2} \lambda^B A^B_\mu \psi$ is the covariant derivative of the spinor field; 
$A^B_\mu $ is the $SU(3)$ gauge potential; $\psi$ is the triplet of the spinor field describing quarks; 
$
	F^B_{\mu \nu} = \partial_\mu A^B_\nu - \partial_\nu A^B_\mu +
	g f_{B C D} A^C_\mu A^D_\nu
$ is the field strength tensor for the $SU(3)$ gauge field, where $f_{BCD}$ are the $SU(3)$ structure constants; $g$ is the coupling constant; $\gamma^\mu$ are the Dirac matrices in the standard representation; $\lambda^B$ are the Gell-Mann indices. For convenience, in what follows we omit the hat notation $\hat{()}$ for all field operators. 

Since such operator equations cannot be solved analytically, they are solved by representing them as an infinite set of equations for the Green's functions (Dayson--Schwinger equations). In the first equation, which is derived by quantum averaging the equation~\eqref{oprtr_YM}, we have a differential equation for
$\left\langle A^B_\mu\right\rangle $. But in this case the resulting equation contains higher-order Green's functions for which one has to write the corresponding equations, and so on, {\it ad infinitum}. 
The same is true for the operator Dirac equation~\eqref{oprtr_D}. As a result, we will have an infinite set of equations for all Green's functions. This process has been discussed in Refs.~\cite{Bender:1999ek,Frasca:2019ysi,Dzhunushaliev:2022apb}. 
In perturbative quantum field theory (for example, in QED), these Dyson--Schwinger-type equations can be solved using Feynman diagrams. However, in a nonperturbative quantum field theory, such as QCD, this approach is not applicable. Let us consider the infinite system of nonperturbative Dyson--Schwinger equations for QCD:
\begin{align}
	\left\langle Q \right| F^{A \mu \nu}_{; \nu} \left| Q \right\rangle
	& = -  \left\langle Q \right| j^{A \mu} \left| Q \right\rangle ,
\label{NP_10}\\
	\left\langle Q \right| A_\rho^B  
		 F^{A \mu \nu}_{; \nu}
	\left| Q \right\rangle & = -  \left\langle Q \right| 
		A_\rho^B j^{A \mu} 
	\left| Q \right\rangle  ,
\label{NP_20}\\
	\left\langle Q \right| 
		 D_\mu \psi -m \psi 
	\left| Q \right\rangle & = 0 ,
\label{NP_30}\\
	\ldots & = \ldots\, ,
\label{NP_40}
\end{align}
where $\left. \left. \right| Q \right\rangle$ is a quantum state describing the given physical system. For convenience, we will use the following notation throughout: 
$	\left\langle Q \right| \cdots \left| Q \right\rangle = \left\langle \cdots \right\rangle$. We will consider static systems in which all observables are time-independent, which considerably simplifies the properties of the field operators. 

\section{Vacuum approximation for the Yang--Mills field}
\label{vacuum_one_field}

In this section, we consider the vacuum, i.e., a physical state in which the quantum expectation value of the components of the gauge potential vanishes. We assume that, in this case, the two-point Green's function $\expval{A^B_\mu A^C_\nu}$ can be represented as a product of one or two scalar functions (factorization). We also assume that the four-point Green's function can be approximately represented as a bilinear combination of two-point Green's functions plus a two-point Green's function multiplied by a certain coefficient, which we refer to as the closure constant.

It is natural to regard the vacuum as a homogeneous and isotropic medium. Therefore, the expectation values of the color electric and magnetic fields must vanish in order to preserve the absence of any preferred direction. Hence, the vacuum expectation values of the color electric and magnetic field vectors should be zero,
\begin{align}
	\expval{E^B_m} & = \expval{F^B_{0 m}} = 
	\expval{\partial_0 A^B_m - \partial_m A^B_0 + g f^{BCD} A^C_0 A^D_m } = 0 , 
\label{5_C_10}\\
	\expval{H^B_m} & = \epsilon_{mnp} \expval{F^B_{np} } = 
	\epsilon_{mnp} \expval{\partial_n A^B_p - \partial_p A^B_n +g f^{BCD} A^C_n A^D_p} , 
\label{5_C_20}
\end{align}
where $m,n,p$ are spatial indices. To understand the appropriate form of this approximation, we recall that we are considering a physical situation in which $\expval{F^B_{\mu\nu}}=0$. The condition $\expval{F^B_{\mu\nu}}=0$ implies that, in a certain gauge, $\expval{A^B_\mu}=0$. We must also choose an approximation for $\expval{A^C_\mu A^D_\nu}$ such that
$f^{BCD}\expval{A^C_\mu A^D_\nu}=0$.
This leads to the conclusion that
$\expval{A^C_\mu A^D_\nu}=\expval{A^D_\mu A^C_\nu}$.
We emphasize that this relation holds only for a static physical system.

\subsection{$\expval{A^B_\mu}= 0$, bilinear combinations $A^B_\mu$ are nonzero, 
a single scalar field for describing $\expval{A^B_\mu A^C_\nu}$}
\label{one_field}

In this subsection, we consider the case in which the gauge potential $A^B_\mu$ in QCD has a vanishing quantum expectation value
\begin{equation}
	\left\langle A^B_\mu \right\rangle \approx 0 . 
\label{scalar_appr}
\end{equation}
The main idea in this case is that, although the quantum-field-theoretic expectation value of the gauge potential $A^B_\mu$ vanishes, the bilinear combination of these field operators is nevertheless nonzero: 
\begin{equation}
	\left\langle A^B_\mu A^C_\nu \right\rangle \neq 0 , 
\label{bilinear}
\end{equation}
and we assume that all even Green's functions $G_{2,4}$ can be described in terms of a scalar function $\phi$ as follows: 
\begin{equation}
	\left\langle A^B_\mu(y) A^C_\nu(x) \right\rangle \approx 
	\zeta^{B C}_{\phantom{B C}\mu \nu} \phi(y) \phi(x) .
\label{3_B_10}
\end{equation}
Physically, this means that our approximation is based on the assumption that the correlation of quantum fluctuations of the gauge potential $A^B_\mu$ at two points $x$ and $y$ can be represented in terms of a single scalar function $\phi$. Here we should recall that we restrict our consideration to the nonperturbative quantization of a strongly nonlinear field in the \emph{static} case. In this case, the properties of the operators of a strongly nonlinear field should differ substantially from those of operators describing noninteracting or weakly interacting fields. In particular, the Green's functions of noninteracting or weakly interacting fields should vanish outside the light cone, whereas this is not necessarily the case for strongly interacting \emph{ } fields. We discuss this issue in greater detail in the concluding part of this section. 

To truncate the infinite system of nonperturbative Dyson--Schwinger equations, we employ the following approximation for the four-point Green's function, schematically written as
$
G_4 \approx G_2^2 + \lambda G_2,
$
i.e., the four-point Green's function is represented as a bilinear combination of two-point Green's functions. We note the appearance of the dimensionful closure constant $\lambda$. Furthermore, we assume that all odd Green's functions vanish,
$
G_{1,3} \approx 0.
$ 

To implement this approximation, we consider the contraction of the Yang--Mills operator equation \eqref{NP_20}
\begin{equation}
	\left\langle A_\mu^B \left( y \right) 
		 F^{B \mu \nu}_{; \nu} \left( x \right)
	\right\rangle = -  \left\langle 
		\bar{\psi} \left( x \right) A_\nu^B\left( y \right) \lambda^B \gamma^\nu \psi \left( x \right) 
	\right\rangle . 
\label{YM_scalar_appr}
\end{equation}
For convenience, all approximations and lengthy calculations for this case are presented in Appendix~\ref{one_scalar_field}. Ultimately, we obtain the following equation for the scalar function $\phi$: 
\begin{equation}
	\left[ 
		\lambda_{\mu \nu} \partial^\mu \partial^\nu \phi - \lambda \Box \phi 
		+ \Lambda \left( \phi^2 - \phi_0 \right) \phi 
	\right] \phi(y)= - j_0  
	\left\langle 
		\bar{\psi}\left( x \right) A^B_\mu(y) \gamma^\mu \lambda^B \psi\left( x \right) 
	\right\rangle ,
\label{YM_scalar_appr_final}
\end{equation}
where $\lambda_{\mu \nu}$ and $\phi_0$ are some constants defined in Appendix~\ref{one_scalar_field}. 

The final question arising in this approximation is which approximation should be chosen for the right-hand side of Eqs.~\eqref{YM_scalar_appr_final}. Clearly, this expression cannot vanish, since the operators $A^B_\mu(y)$ and $\bar{\psi}(x) \psi(x)$ are coupled through the operator equations \eqref{NP_10} and \eqref{NP_30}. We address this issue in Sec.~\ref{one_field}. 

In brief (with the details given in Sec.~\ref{current_appr}), the approximation for the right-hand side of Eq.~\eqref{YM_scalar_appr_final} takes the following form: 
\begin{equation}
	\left\langle 
		\bar{\psi} \left( x \right) A_\nu^B\left( y \right) \lambda^B \gamma^\nu \psi \left( x \right) 
	\right\rangle \approx j_0 \phi(y) 
	\bar{\varsigma}\left( x \right) \varsigma \left( x \right) . 
\label{current_appr_in}
\end{equation}
The physical meaning of this approximation is that the correlation
$
\left\langle
\bar{\psi} \left( x \right) A_\nu^B\left( y \right) \lambda^B \gamma^\nu \psi \left( x \right)
\right\rangle
$
of the quantum fluctuations of the operators $A^B_\mu(y)$ and $\bar{\psi}(x) \psi(x)$ can be factorized in terms of the scalar function $\phi$, which describes the expectation value of $A^B_\mu A^{B\mu}$, and the spinor function $\varsigma$, which provides an approximation for the vacuum condensate 
$\expval{\bar{\psi}(x) \psi(x)}$.

All of this allows us to rewrite Eq.~\eqref{YM_scalar_appr} in the following form:
\begin{equation}
	\lambda_{\mu \nu} \partial^\mu \partial^\nu \phi - \lambda \Box \phi
	+ \Lambda \left( \phi^2 - \phi_0 \right) \phi
	= - j_0 \bar{\varsigma}\left( x \right) \varsigma \left( x \right) ,
\label{YM_scalar_appr_final_2}
\end{equation}
where we have divided by $\phi(y)$.

Thus, in this section we have obtained Eq.~\eqref{YM_scalar_appr_final_2} for the scalar field $\phi$, which describes the condensate
$
\left\langle \hat{A}^B_\mu \hat{A}^{B \mu} \right\rangle,
$
with the vacuum quark condensate
$
\left\langle \hat{\bar{\psi}} \hat{\psi} \right\rangle
$
acting as its source.

The approximation for Green's functions involving the product of $A^B_\mu$ and $\psi$ will be discussed in greater detail in Sec.~\ref{quark_approx}. 

\subsection{$\left\langle \hat{A}^B_\mu\right\rangle = 0$, bilinear combinations $\hat{A}^B_\mu$ are nonzero, two scalar fields}
\label{two_condensates}

In this subsection, we consider a situation in which two scalar fields can be used to describe the physical system under consideration. This may occur when one part of the degrees of freedom, $A^a_\mu \in G \subset SU(3)$, with $\left\langle \hat{A}^a_\mu \right\rangle = 0$, is described by one scalar field $\phi$, while the remaining degrees of freedom, $A^m_\mu \in SU(3)/G$, with $\left\langle \hat{A}^m_\mu \right\rangle = 0$, are described by another scalar field $\chi$.

In this case, we have to consider two contractions of Eq.~\eqref{NP_20}: 
\begin{align}
	\left\langle A_\nu^a \left( y \right) 
		 F^{a \nu \mu}_{; \mu} \left( x \right)
	\right\rangle = -  \left\langle 
		\bar{\psi}\left( x \right) A_\nu^a \left( y \right) \lambda^a \gamma^\nu \psi\left( x \right) 
	\right\rangle , 
\label{phi_eqn}\\
	\left\langle A_\nu^m \left( y \right) 
		 F^{m \nu \mu}_{; \mu} \left( x \right)
	\right\rangle = -  \left\langle 
		\bar{\psi}\left( x \right) A_\nu^m \left( y \right) \lambda^m \gamma^\nu \psi\left( x \right) 
	\right\rangle . 
\label{chi_eqn}
\end{align}
Here 
\begin{eqnarray}
	F^a_{\mu \nu} = 
		\mathcal{F}^a_{\mu \nu} + g f^{a m n} A^m_\mu A^n_\nu 
	=  \mathcal{F}^a_{\mu \nu} + \mathfrak{F}^a_{\mu \nu} , 
\label{field_strength_tensor_1}\\
	F^m_{\mu \nu} = 
		\partial_\mu A^m_\nu - \partial_\nu A^m_\mu 
		+ g f^{m n a} \left( A^n_\mu A^a_\nu - A^a_\mu A^n_\nu\right) ,
\label{field_strength_tensor_2}
\end{eqnarray}
where 
$
	\mathcal{F}^a_{\mu \nu} = 
	\partial_\mu A^a_\nu - \partial_\nu A^a_\mu +
	g f_{a b c} A^b_\mu A^c_\nu 
$ is the strength tensor for $A^a_\mu \in G$, 
$
	\mathfrak{F}^a_{\mu \nu} = f^{a m n} A^m_\mu A^n_\nu 
$, and $A^m_\mu \in SU(3) / G$.

After carrying out the lengthy calculations presented in Appendix~\ref{two_scalar_fields}, we obtain the following equations for the scalar functions $\phi$ and $\chi$, which describe the Green's functions for the components 
$A^a_\mu \in G \subset SU(3)$ and $A^m_\mu \in SU(3) / G$:
\begin{align}
	\phi(y) \left[ 
		\lambda_{\mu \nu} \partial^\mu \partial^\nu \phi - \lambda \Box \phi 
		+ \left( 
			\lambda_1 \phi^2 + \lambda_2 \chi^2 + \lambda_3
		\right) \phi \right] & = 
	- \left\langle 
		\bar{\psi}\left( x \right) A_\mu^a \left( y \right) \lambda^a \gamma^\mu \psi\left( x \right) 
	\right\rangle , 
\label{3_C_10}\\
	\chi(y) \left[ 
			\xi_{\mu \nu} \partial^\nu \partial^\mu \chi - \xi \Box \chi 
		+ \left( 
			\xi_1 \chi^2 + \xi_2 \phi^2 + \xi_3
		\right) \chi 
		\right] & = -  \left\langle 
		\bar{\psi}\left( x \right) A_\mu^m \left( y \right) \lambda^m \gamma^\mu \psi\left( x \right) 
	\right\rangle . 
\label{3_C_20}
\end{align}

A highly nontrivial issue is the choice of an appropriate approximation for the right-hand sides of Eqs.~\eqref{3_C_10} and \eqref{3_C_20}. This issue will be considered in Sec.~\ref{current_appr}. Using the results of that section, we obtain the following approximation for the right-hand sides of Eqs.~\eqref{3_C_10} and \eqref{3_C_20}:
\begin{align}
	\left\langle 
		\bar{\psi}\left( x \right) A_\mu^a \left( y \right) \lambda^a \gamma^\mu \psi\left( x \right) 
	\right\rangle 
	& \approx j_1 \phi(y) \bar{\varsigma}\left( x \right) \varsigma \left( x \right) , 
\label{3_C_30}\\
	\left\langle 
		\bar{\psi}\left( x \right) A_\mu^m \left( y \right) \lambda^m \gamma^\mu \psi\left( x \right) 
	\right\rangle &  \approx j_2 \chi(y) \bar{\varsigma}\left( x \right) \varsigma \left( x \right) , 
\label{3_C_40}
\end{align}
which ultimately yields the following system of equations for the condensates $\phi$ and $\chi$:
\begin{align}
	\lambda_{\mu \nu} \partial^\mu \partial^\nu \phi - \lambda \Box \phi 
	+ \left( 
				\lambda_1 \phi^2 + \lambda_2 \chi^2 + \lambda_3
			\right) \phi & = - j_1 \bar{\varsigma}\left( x \right) \varsigma \left( x \right) , 
\label{3_C_50}\\
	\xi_{\mu \nu} \partial^\nu \partial^\mu \chi - \xi \Box \chi 
	+ \left( 
			\xi_1 \chi^2 + \xi_2 \phi^2 + \xi_3
	\right) \chi  & = - j_2 \bar{\varsigma}\left( x \right) \varsigma \left( x \right) , 
\label{3_C_60}
\end{align}
where we have divided by $\phi(y), \chi(y)$. 

Thus, in this section, we have obtained Eqs.~\eqref{3_C_50} and \eqref{3_C_60} for the scalar fields $\phi$ and $\chi$, which describe the condensates
$
\left\langle A^{a, m}_\mu A^{b,n}_\nu \right\rangle,
$
whose sources are the condensates
$	
	\expval{\bar{\psi}(x) A_\mu^{a, m} (y) \lambda^{a, m} \gamma^\mu \psi (x)}
$. 

\subsection{Physical assumptions for the vacuum case}
\label{phys_assumptions}

For clarity, let us summarize the physical assumptions used in Sec.~\ref{vacuum_one_field} to obtain the first approximation to the infinite system of nonperturbative Dyson--Schwinger equations for the vacuum:
\begin{itemize}
\item We consider a static problem. This means that the correlation of quantum fluctuations of strongly interacting fields at spatially separated points can be nonzero even when the points are separated by a spacelike interval. This fundamentally distinguishes the situation considered here from the corresponding case of weakly interacting fields, for which such correlations vanish outside the light cone. Physically, this can be understood as follows. In the case of weakly interacting fields (for example, in QCD), the correlation is mediated by quanta propagating at a speed lower than the speed of light, whereas in the case of strongly interacting fields, the correlation is described by the corresponding nonlinear equation, which admits time-independent solutions. By comparison with classical field theory, linear equations (for example, Maxwell's equations) possess wave solutions propagating at the speed of light, whereas nonlinear equations (for example, the Yang--Mills equations) admit nontrivial static solutions, such as the ’t Hooft--Polyakov monopole.
\item Factorization: the quantum-field-theoretic expectation value of the product of two gauge-field operators \emph{in the static case} can be represented as a product of either one or two scalar functions evaluated at the corresponding points.
\item An essential feature of our model is that the field components $A^B_\mu(x)$ and $A^C_\nu(y)$ at spatially separated points $x$ and $y$ can be correlated even when they have different color indices $B,C$ as well as different spacetime indices $\mu,\nu$. This is due to the fact that all these components are coupled through the nonlinear Yang--Mills equation.
\item The truncation of the infinite system of nonperturbative Dyson--Schwinger equations is performed by representing four-point Green's functions as bilinear combinations of two-point Green's functions, which leads to the appearance of new dimensionful constants, referred to as ``closure constants.''
\item Within the approximation considered, the odd Green's functions of the Dyson--Schwinger system vanish, whereas the even Green's functions are nonzero.
\end{itemize}

\section{Approximation for ``almost classical'' and ``almost vacuum'' degrees of freedom
}
\label{two_fields}
In this section, we consider a situation in which the gauge potential in QCD has ``almost classical'' and ``almost vacuum''  degrees of freedom. By ``almost classical'' we mean that their quantum expectation values are nonzero, whereas ``almost vacuum'' means that their quantum expectation values are much smaller than the characteristic values of the ``almost classical'' values:
\begin{align}
	\left\langle \hat{A}^a_\mu \right\rangle \neq 0, 
\label{3_D_10}\\
	\left\langle \hat{A}^m_\mu \right\rangle \approx 0 . 
\label{3_D_20}	
\end{align}
In this section, we restore the hat notation $\hat{}$ for all operators. The most natural assumption concerning the indices $a$ and $m$ is the following:
$\hat{A}^a_\mu \in G \subset SU(3)$ and $\hat{A}^m_\mu \in SU(3) / G$, where $G$ is a subgroup of $SU(3)$. For example,
$G = SU(2) \times U(1)$.

\subsection{Color/spatial asymmetry}
\label{color_asymm}

Consider the expression
$
\expval{f^{A B C} \hat{A}^B_\mu \left( x_1\right) \hat{A}^C_\nu \left( x_2 \right) }
$
for $x_2 \neq x_1$. Due to the antisymmetry of the structure constants $f^{A B C}$, this expression vanishes unless there is some asymmetry of the operators $\hat{A}^B_\mu$ either in color space with respect to the indices $B,C$ or in four-dimensional spacetime with respect to the indices $\mu,\nu$. Such a situation may arise, for example, if the physical system possesses preferred directions generated by color electric or magnetic fields between quarks. These may be, for instance, a longitudinal color electric field in a flux tube connecting a quark and an antiquark, or a $Y$-shaped distribution of such a field between three quarks in a hadron. This means that there are field lines of the averaged field
$
\expval{\hat{E}^A_\mu \left( x \right)}
$
with some specific value or values of the color index $A$ or of the spacetime index $\mu$, which corresponds to an asymmetry either in color space or in four-dimensional Minkowski spacetime.

Analogous considerations apply to the four-point Green's function as well, 
\begin{align}
	& G_4 \left( x_1, x_2, x_3, x_4\right) =  f^{A B C} f^{M N P} 
		\expval{\hat{A}^B_\mu \left( x_1\right) \hat{A}^C_\nu \left( x_2 \right)  
		\hat{A}^N_\rho \left( x_3\right) \hat{A}^P_\sigma \left( x_4 \right) }
\nonumber \\
	& = 
	\frac{1}{4} f^{A B C} f^{M N P} 
	\expval{ 
		\hat{A}^B_\mu \left( x_1\right) \hat{A}^C_\nu \left( x_2 \right)	
		- \hat{A}^C_\mu \left( x_1\right) \hat{A}^B_\nu \left( x_2 \right)	
	}
	\expval{\hat{A}^M_\mu \left( x_3\right) \hat{A}^N_\nu \left( x_4 \right)	
		- \hat{A}^N_\mu \left( x_3\right) \hat{A}^M_\nu \left( x_4 \right)	}
	\neq 0 , 
\label{Green_4}
\end{align}
which is nonzero if there is an asymmetry either in color space or in Minkowski spacetime.


\subsection{Approximation for asymmetric configurations
}
\label{asymmetry}
According to the above assumption, we have ``almost classical'' degrees of freedom, for which the corresponding equation takes the form 
\begin{equation}
	\left\langle \hat F^{a \mu \nu}_{; \nu} \right\rangle
	= -  \left\langle \hat j^{a \mu} \right\rangle . 
\label{3_D_30}
\end{equation}
We aim to truncate all the remaining equations and show that, in this case, the resulting equation is the field equation of Proca-type theory with sources describing quarks and the gluon condensate arising from the ``almost vacuum'' degrees of freedom. We introduce the averaged field-strength tensors $F^{a, m}_{\mu \nu}$,
\begin{eqnarray}
	F^a_{\mu \nu} = \expval{\hat{F}^a_{\mu \nu}} = 
	\mathcal{F}^a_{\mu \nu} + g f^{a m n} \left\langle 
		\hat{A}^m_\mu \hat{A}^n_\nu 
	\right\rangle 
	=  \mathcal{F}^a_{\mu \nu} + \mathfrak{F}^a_{\mu \nu} , 
\label{3_D_40}\\
	\hat{F}^m_{\mu \nu} = 
	\partial_\mu \hat{A}^m_\nu - \partial_\nu \hat{A}^m_\mu 
		+ g f^{m n a} \left( \hat{A}^n_\mu A^a_\nu - A^a_\mu \hat{A}^n_\nu\right) ,
\label{3_D_50}
\end{eqnarray}
where 
$
	\mathcal{F}^a_{\mu \nu} = 
	\partial_\mu A^a_\nu - \partial_\nu A^a_\mu +
	g f_{a b c} A^b_\mu A^c_\nu 
$, $A^a_\mu \approx \expval{\hat{A}^a_\mu}$ and it is a strength tensor for ``almost classical'' degrees of freedom $A^a_\mu \in G$, 
$
	\mathfrak{F}^a_{\mu \nu} = f^{a m n} \left\langle \hat{A}^m_\mu \hat{A}^n_\nu \right\rangle 
$, 
and $\hat{A}^m_\mu \in SU(3) / G$ are ``purely quantum'' degrees of freedom. Substituting  \eqref{3_D_40} in \eqref{3_D_30}, we have the following result: 
\begin{equation}
	\widetilde{D}_\nu \left( 
	\mathcal{F}^{a \mu \nu} + \mathfrak{F}^{a \mu \nu} 
	\right) 
	+ g f^{a m n} \left\langle \hat{A}^m_\nu \hat{F}^{n \mu \nu} \right\rangle 
	= -  \left\langle \hat j^{a \mu} \right\rangle ,
\label{3_D_60}
\end{equation}
where 
$
	 \widetilde{D}_\nu \mathcal{F}^{a \mu \nu} 
	 	= \partial_\nu \mathcal{F}^{a \mu \nu} 
	 	+ g f^{a b c} A^b_\nu \mathcal{F}^{c \mu \nu} 
$ is a gauge derivative in subgroup $G \subset U(3)$, and the same is true for $\mathfrak{F}^{a \mu \nu}$. 

For clarity of presentation, the intermediate calculations of the left-hand side of Eq.~\eqref{3_D_60} are given in Appendix~\ref{nonzero_zero}, leading to the following result: 
\begin{equation}
	\widetilde{D}_\nu	\mathcal{F}^{a \mu \nu} 
	+ g \widetilde{D}_\nu \left( \zeta^{a \mu \nu} \chi^2 \right) 
	- \frac{g}{2} \widetilde{D}_\nu \left( \zeta^{a \nu \mu} \chi^2 \right) 
	- \frac{g}{2} \widetilde{D}^\mu \left( \zeta^{a} \chi^2 \right) 
	+ \xi^{a b \mu \nu} A^b_\nu \chi^2 
	= - \left\langle \hat{\bar{\psi}} \gamma^\mu \lambda^a \hat{\psi} \right\rangle , 
\label{3_D_75}
\end{equation}
where the scalar field $\chi$ describes the condensate $\expval{A^m_\mu A^n_\nu}$. To obtain a closed system of equations, it is necessary to supplement it with an equation for the ``almost quantum'' degrees of freedom. To this end, we use the contraction
\begin{equation}
	\expval{\hat{A}^m_\mu(y) D_\nu \hat{F}^{m \mu \nu}} = 
	- \expval{\hat{A}^m_\mu(y) \hat{j}^{m \mu}} . 
\label{3_D_80}
\end{equation}
The detailed calculations of the left-hand side of Eq.~\eqref{3_D_80} are given in Appendix~\ref{nonzero_zero}, ultimately yielding the following equation:
\begin{equation}
\begin{split}
& \chi(y) \left[ 
		\zeta^{\mu \nu} \partial_\mu \partial_\nu \chi 
		+ g \zeta^{a \mu \nu} \mathcal{F}^a_{\mu \nu} \chi 
		+ g \nabla_\nu 
		\left( 
			 \zeta^a A^{a \nu} \chi - \zeta^{a \mu \nu} A^a_\mu \chi  
		\right) 
		\right. 
\\
	& \left. 
	- g 
	\left( 
		\zeta^{a \mu \nu} A^a_\nu - \zeta^a A^{a \mu} 
		\right) \nabla_\mu \chi 
		+ \Lambda \left( 
			\chi^2 - \chi_0
	\right) \chi 
	- m^{ab \mu \nu} A^a_\nu A^b_\mu \chi 
	\right] = - \expval{\hat{A}^m_\mu(y) \hat{\bar{\psi}} \gamma^\mu \lambda^m \hat{\psi}} . 
\end{split}
\label{3_D_90}
\end{equation}

Ultimately, we obtain the following system of equations describing the situation with
$\hat{A}^a_\mu \neq 0$ and $\hat{A}^m_\mu = 0$:
\begin{align}
	\widetilde{D}_\nu	\mathcal{F}^{a \mu \nu} 
	+ g \widetilde{D}_\nu \left( \zeta^{a \mu \nu} \chi^2 \right) 
	- \frac{g}{2} \widetilde{D}_\nu \left( \zeta^{a \nu \mu} \chi^2 \right) 
	- \frac{g}{2} \widetilde{D}^\mu \left( \zeta^{a} \chi^2 \right) 
	+ \xi^{a b \mu \nu} A^b_\nu \chi^2 
	& = - \expval{\hat{\bar{\psi}} \gamma^\mu \lambda^a \hat{\psi}}, 
\label{3_D_100}\\
	\chi(y) \left[ 
		\zeta^{\mu \nu} \partial_\mu \partial_\nu \chi 
		+ g \zeta^{a \mu \nu} \mathcal{F}^a_{\mu \nu} \chi 
		+ g \nabla_\nu 
		\left( 
			 \zeta^a A^{a \nu} \chi - \zeta^{a \mu \nu} A^a_\mu \chi  
		\right) 
	\right. & 
\nonumber \\
	\left. 
		- g 
		\left( 
			\zeta^{a \mu \nu} A^a_\nu - \zeta^a A^{a \mu} 
			\right) \nabla_\mu \chi 
			+ \Lambda \left( 
				\chi^2 - \chi_0
		\right) \chi 
		- m^{ab \mu \nu} A^a_\nu A^b_\mu \chi 
	\right] &= - \expval{\hat{A}^m_\mu(y) \hat{\bar{\psi}} \gamma^\mu \lambda^m \hat{\psi}} . 
\label{3_D_110}
\end{align}
A highly nontrivial and challenging issue is the choice of an appropriate approximation for the right-hand sides of these equations: $\left\langle \hat{\bar{\psi}} \gamma^\mu \lambda^a \hat{\psi} \right\rangle$ and 
$\expval{\hat{A}^m_\mu(y) \hat{\bar{\psi}} \gamma^\mu \lambda^m \hat{\psi}}$. Approximation for the right-hand sides of
 \eqref{3_D_100} and \eqref{3_D_110} will be given in Appendix~\ref{Dirac_eqn}, see Eqs.~\eqref{В_130} and \eqref{В_140}. 

\subsection{Physical assumptions for ``almost classical'' and ``almost vacuum'' degrees of freedom
}
\label{phys_assumptions_2}
In addition to the assumptions listed in Sec.~\ref{phys_assumptions}, we make the following assumption:
In the presence of color/spatial asymmetry, all degrees of freedom can be divided into ``almost classical''  degrees of freedom $A^a_\mu \in G \subset SU(3)$ such that $\expval{\hat{A}^a_\mu} \approx A^a_\mu$ and ``almost vacuum'' degrees of freedom $A^m_\mu \in SU(3)/G$ such that $\expval{\hat{A}^m_\mu} \approx 0$.

\section{Approximation for a quark field}
\label{quark_approx}
In this section, we propose an approximation for the fermions entering the Dirac equation, as well as for the right-hand sides of the Yang--Mills equations, where the fermions act as sources of the gauge field. Thus, our task is to obtain approximations for the Dirac equation \eqref{NP_30} and for the right-hand sides of Eqs.~\eqref{YM_scalar_appr_final}, \eqref{3_C_10}, \eqref{3_C_20}, \eqref{3_D_100}, and \eqref{3_D_110}:
\begin{align}
	\expval{
	\frac{\imath}{2} 
	\qty(\gamma^\mu)_{\alpha \beta}
	\qty(\nabla_\mu)_{ij} \psi_{\beta j} - m \psi_{\alpha i} } & = 0, 
\label{5_10}\\ 
	\expval{\bar{\psi}\left( y \right) A^B_\mu(x) \gamma^\mu \lambda^B \psi\left( x \right) } & \approx ? 
\label{5_20}\\
	\expval{\hat{A}^m_\mu(y) \hat{\bar{\psi}}(x) \gamma^\mu \lambda^m \hat{\psi}(x)} & \approx ? 
\label{5_23}\\
	\expval{\bar{\psi}(x) \gamma^\mu \lambda^B \psi(x)} & \approx ? 
\label{5_25}
\end{align}
Here the spinor describing quarks has two indices: $\psi_{\alpha i}$, where $\alpha = 1, \ldots 4$ is the spinor index, and $i = 1, 2, 3 = red, green ,blue$ is the color index, and 
$\qty(\nabla_\mu)_{ij} = \delta_{i j} \partial_\mu + \frac{\imath}{g} \lambda^a_{i j} A^a_\mu$. 
Here, the \textit{very} challenging problem is to determine an appropriate approximation for an expression involving a product of spinor and gauge fields: 
$
	\left\langle 
		\bar{\psi} \lambda^B A^B_\mu \gamma^\mu \psi 
	\right\rangle 
$. 

In the following subsections, we consider separately the vacuum case and the case of asymmetric color/spatial configurations of the quantum field.

\subsection{Vacuum case}
\label{vacuum_quark}

The main problem here is to find an appropriate approximation for the three-point Green's functions involving the product of $\hat{\bar{\psi}}$, $\hat{A}^B_\mu$, and $\hat{\psi}$; see Eqs.~\eqref{5_20}--\eqref{5_25}. The difficulty is that these three-point Green's functions cannot be factorized, as was done in Eqs.~\eqref{app_A_20} and \eqref{app_B_30}, into a product of $\expval{\hat{A}^B_\mu}$ and $\expval{\hat{\bar{\psi}} \hat \psi}$. In the vacuum case, all the quantities $\hat{A}^B_\mu$ and $\hat \psi$ have vanishing expectation values; however, since they are coupled through the Dirac and Yang--Mills equations, the three-point Green's functions \eqref{5_20}--\eqref{5_25} do not vanish. A physically motivated and intuitive approximation is to factorize these Green's functions as products 
of quantities depending on the two-point Green's functions of the quark fields and the gauge potential.

\subsubsection{Approximation for the Dirac equation}
\label{Dirac_appr}

Let us begin by seeking an approximation for the Dirac equation \eqref{5_10}. We assume that, in the vacuum case, the triplet of fermion fields $\psi_{\alpha i}$ can be approximately described in terms of the singlet fermion field $\varsigma_{\alpha}$. To this end, we consider the expectation value of the Dirac equation multiplied by a component of the Dirac-conjugate spinor evaluated at the point $y$:
\begin{equation}
	\expval{\hat{\bar{\psi}}_{\gamma i}(y)
	\left[ 
		\imath
		\qty(\gamma^\mu)_{\alpha \beta}
		\qty(D_\mu)_{ij} \hat{\psi}_{\beta j} - m \hat{\psi}_{\alpha i}
	\right]  
	} = 0 .
\label{5_a_10}
\end{equation}

Let us now consider an appropriate approximation for quadratic combinations of the quark field
$
	\expval{\hat{\bar{\psi}}_{\gamma k}(y) \qty(\gamma^\mu)_{\alpha \beta}
	\partial_\mu \hat{\psi}_{\beta i} }
$, 
$
	\expval{\hat{\bar{\psi}}_{\gamma k}(y) \hat{\psi}_{\alpha i}}
$. 
In this case
\begin{align}
	\expval{\hat{\bar{\psi}}_{\gamma i}(y) \hat{\psi}_{\alpha i}} & \approx 
	\bar{\varsigma}_{\gamma}(y) \varsigma_{\alpha} , 
\label{5_a_30}\\
	\expval{\hat{\bar{\psi}}_{\gamma i}(y) \qty(\gamma^\mu)_{\alpha \beta}	\partial_\mu \hat{\psi}_{\beta i} } 
	& \approx \bar{\varsigma}_{\gamma}(y) \qty(\gamma^\mu)_{\alpha \beta}	
	\partial_\mu \varsigma_{\beta}. 
\label{5_a_20}
\end{align}
We emphasize that the spinor $\varsigma \neq \psi$. The physical meaning of introducing this spinor is that it \emph{approximately} describes expectation value of quadratic combinations of the spinor $\psi$. Let us recall that, in the Introduction, we proposed the following rule: \textit{the approximate description of bosonic fields with integer spin (in our case, gauge fields) is achieved by introducing bosonic fields (the scalar fields $\phi, \chi$ and/or the vector field $A^a_\mu \in G \subset SU(3)$); the approximate description of fermions is achieved by means of a spinor field.} We have now implemented this prescription by introducing the singlet spinor $\varsigma$ to approximately describe the quarks $\psi$,  with $\psi$ being a triplet.

The next step is to seek an approximation for the expression 
$
	\expval{\hat{\bar{\psi}}_{\gamma i}(y) A^B_\mu(x) \left(\gamma^\mu\right)_{\alpha\beta}
	\left(\lambda^B\right)_{ij}	\hat{\psi}_{\beta j}(x)}
$. In this expression, the expectation values of all fields vanish, $\expval{A^B_\mu} = 0$ and $\expval{\hat{\psi}} = 0$, which means that we are dealing with sea quarks and the vacuum gauge field. However, the vacuum expectation value of this product cannot vanish, since all these fields are coupled through the Dirac and Yang-Mills equations. The nonzero quantities in this case are the two-point Green's functions
$\expval{\hat{\bar{\psi}}_{\alpha i}(y) \hat{\psi}_{\beta i}(x)}$ and $\expval{A^B_\mu A^{B \mu}}$.
According to Appendices~\ref{one_scalar_field} and \ref{two_scalar_fields}, the two-point Green's function can be approximately represented in the form
\begin{equation}
	\expval{A^B_\mu A^{B \mu}} \approx \begin{cases}
		\zeta^{B B}_{\phantom{B B}\mu \nu} \phi^2 , &\text{for the case from Sec.  \ref{one_field}}\\
		\zeta^{b b}_{\phantom{b c}\mu \nu} \phi^2 + 
		\zeta^{m m}_{\phantom{m n}\mu \nu} \chi^2 , 
		&\text{for the case from Sec.  \ref{two_condensates}} . 
	\end{cases}
\label{Green_2}
\end{equation}
This allows us to propose the following approximation:
\begin{equation}
	\expval{\hat{\bar{\psi}}_{\gamma i}(y) A^B_\mu \left(\gamma^\mu\right)_{\alpha\beta}
	\left(\lambda^B\right)_{ij}	\hat{\psi}_{\beta j}} \approx 
	\begin{cases}
		\delta_1 \phi \bar{\varsigma}_\gamma(y) \varsigma_\alpha 
		+ \delta_3 \bar{\varsigma}_\gamma(y) \varsigma_\beta 
		\left( 
			\bar{\varsigma}_\alpha \varsigma_\alpha
		\right)  , &\text{for the case from Sec.  \ref{one_field}}\\
		\left( \delta_1 \phi 	+ \delta_2 \chi \right) \bar{\varsigma}_\gamma(y) \varsigma_\alpha 
		+ \delta_3 \bar{\varsigma}_\gamma(y) \varsigma_\beta 
		\left( 
			\bar{\varsigma}_\alpha \varsigma_\alpha
		\right) , 
		&\text{for the case from Sec.  \ref{two_condensates}} . 
	\end{cases}
\label{5_a_40}
\end{equation}
This expression is essentially the simplest and practically unique Lorentz-invariant expression involving quadratic combinations of the gauge field and the spinor $\varsigma$, \textit{whose physical meaning is that it approximately describes the quark condensate generated by (virtual) sea quarks bound by the vacuum gauge field.}

All of this leads to the following form of the Dirac equation \eqref{5_a_10} in the vacuum case:
\begin{align}
	\imath 
	\qty(\gamma^\mu)_{\alpha \beta}
	\partial_\mu \varsigma_{\beta} 
	+ \delta_1 \phi \varsigma_\alpha 
	+ \delta_3 \varsigma_{\alpha}  \left( 
			\bar{\varsigma}_\beta \varsigma_\beta
		\right) 
	- m \varsigma_{\alpha} & = 0 , \text{for the case from Sec.  \ref{one_field}}
\label{5_a_45}\\
	\imath 
	\qty(\gamma^\mu)_{\alpha \beta}
	\partial_\mu \varsigma_{\beta} 
	+ \left( \delta_1 \phi 	+ \delta_2 \chi \right) \varsigma_\alpha 
	+ \delta_3 \varsigma_{\alpha}  \left( 
			\bar{\varsigma}_\beta \varsigma_\beta
		\right) 
	- m \varsigma_{\alpha} & = 0 , \text{for the case from Sec.  \ref{two_condensates}} 
\label{5_a_47}
\end{align}
where we have divided by $\varsigma(y)$. Let us note the appearance, within this approximation, of the nonlinear term $\left( \bar{\varsigma} \varsigma\right) ^2$, which has a substantial influence on the possible solutions of this equation. In the last century, Refs.~\cite{Finkelstein:1951zz} and \cite{Finkelstein:1956} investigated the nonlinear Dirac equation and showed that it possesses a global minimum in the energy spectrum of its solutions. At that time, the term ``mass gap'' had not yet been introduced, and the authors referred to this state as the ``lightest stable particle.'' Subsequently, Refs.~\cite{Dzhunushaliev:2020qwf} and \cite{Dzhunushaliev:2021apa} showed that this property is preserved in Yang--Mills theory coupled to a nonlinear Dirac equation: a mass gap exists. Therefore, it is reasonable to expect that a mass gap will also exist within our approximation.

\subsubsection{Approximation for
$
	\expval{\hat{\bar{\psi}} (x) \lambda^{a, m} \hat{A}^{a, m}_\mu(y) \gamma^\mu \hat{\psi}(x) }
$}
\label{current_appr}

Consider now approximation for
$
	\expval{\hat{\bar{\psi}} (x) \lambda^{a, m} \hat{A}^{a, m}_\mu(y) \gamma^\mu \hat{\psi}(x) }
$ entering the right-hand sides of Eqs.~\eqref{YM_scalar_appr_final}, \eqref{3_C_10}, \eqref{3_C_20}, and \eqref{3_D_110}. We cannot approximately represent these vacuum expectation values as products of vacuum expectation values $\expval{A^{a, m}(y)}$ and $\expval{\bar{\psi} \lambda^{a, m} \gamma^\mu \psi}$, since in this case the three-point Green's function
$
	\left\langle 
		\bar{\psi}(x) \lambda^{a, m} A^{a,m}_\mu(y) \gamma^\mu \psi(x) 
	\right\rangle = 0 
$, but the vacuum expectation value of this product cannot vanish, since all these fields are coupled through the Dirac and Yang-Mills equations. In this case, the nonzero quantity is the quark condensate
$\expval{\bar{\psi} \psi}$ and the two-point Green's functions
$\expval{A^{a,m}_\mu(y) A^{b,n}_\nu(x)} \neq 0$. According to our approximations \eqref{app_A_20}, \eqref{app_B_30}, \eqref{app_B_70}, 
these Green's functions
$
	\expval{A^{a}_\mu(y) A^{b}_\nu(x)} \approx 
	\zeta^{ab}_{\phantom{ab} \mu \nu} \phi(y) \phi(x) 
$ or
$
	\expval{A^m_\mu(y) A^n_\nu(x)} \approx \zeta^{mn}_{\phantom{mn} \mu \nu} \chi(y) \chi(x)
$. This enables us to suggest the following approximation:
\begin{align}
	\left\langle 
		\bar{\psi}(x) \lambda^a A^a_\mu(y) \gamma^\mu \psi(x) 
	\right\rangle & \approx j_1 \phi(y) \bar{\varsigma}(x) \varsigma(x) ,
\label{5_C_30}\\
	\left\langle 
		\bar{\psi}(x) \lambda^m A^m_\mu(y) \gamma^\mu \psi(x) 
	\right\rangle & \approx j_2 \chi(y) \bar{\varsigma}(x) \varsigma(x) , 
\label{5_C_32}
\end{align}
where $j_{1, 2}$ are some coefficients. 
For the approximation considered in Sec.~\ref{one_field}, we use \eqref{5_C_30}, whereas for the approximation considered in Sec.~\ref{two_condensates}, we use \eqref{5_C_30} and \eqref{5_C_32}.

\subsection{Approximation for asymmetric configurations}
\label{asymm_Dirac} 
 
Consider the expectation value of the Dirac equation
\begin{equation}
	\expval{
	\hat{\bar{\psi}}_{\gamma k}(y) 
	\left[ 
		\imath \qty(\gamma^\mu)_{\alpha \beta}
		\qty(D_\mu)_{ij} \hat{\psi}_{\beta j} - m \hat{\psi}_{\alpha i}
	\right] 
	} = 0 . 
\label{5_B_10}
\end{equation}
Here, the expression in square brackets is a function of $x$, and the covariant derivative of the spinor field is 
$
	D_\mu \psi_{\alpha i}= 
	\partial_\mu \psi_{\alpha i} - \imath \frac{g}{2} \qty(\lambda^B)_{ij} 
	\qty(\gamma^\mu)_{\alpha \beta} A^B_\mu \psi_{\beta j} 
$. We also note that, in contrast to the contraction of the Dirac equation \eqref{5_a_10} in Sec.~\ref{Dirac_appr}, there is no summation over the pair of indices $k,i$ here. We consider asymmetric configurations in which $\expval{\psi} \neq 0$ and there are ``almost classical degrees of freedom'' $A^a_\mu \in G \subset SU(3)$ with nonzero quantum expectation values, $\expval{A^a_\mu} \neq 0$, and ``almost quantum degrees of freedom'' $A^m_\mu \in SU(3) / G$ with vanishing expectation values, $\expval{A^m_\mu} = 0$.

The quark-field operator $\hat{\psi}$ can be represented as the sum of its expectation value $\expval{\hat{\psi}} = \vartheta$ and the quantum fluctuations $\widehat{\delta \psi}$:
$
	\hat{\psi}_{\alpha i} = \vartheta_{\alpha i} + \widehat{\delta \psi}_{\alpha i}.
$
Our main assumption is that the Dirac equation \eqref{5_B_10} can be approximately replaced by an equation for $\vartheta$ and an equation for the two-point Green's function $\expval{\widehat{\delta \bar{\psi}}{\gamma k}(y) \widehat{\delta \psi}_{\alpha i}}$, which is factorized as follows:
$
	\expval{\widehat{\delta \bar{\psi}}_{\gamma k}(y) \widehat{\delta \psi}_{\alpha i}} 
	\approx \bar{\varsigma}_{\gamma k}(y) \varsigma_{\alpha i}
$. 

All lengthy calculations leading to these equations, together with the approximations employed, are presented in Appendix~\ref{Dirac_eqn}, ultimately yielding the following two equations:
\begin{align}
		\imath \qty(\gamma^\mu)_{\alpha \beta}
		\qty(\tilde{D}_\mu)_{ij} \vartheta_{\beta j} - m \vartheta_{\alpha i} & = 0 , 
\label{5_B_12}\\
	\imath \qty(\gamma^\mu)_{\alpha \beta}
	\qty(\tilde{D}_\mu)_{ij} \varsigma_{\beta j} - 
	\left( 
		m - \kappa_1 \chi 
	\right) \varsigma_{\alpha i} + \kappa_2 \left( \bar{\varsigma} \varsigma \right) \varsigma_{\alpha i}  
	& = 0 . 
\label{5_B_14}
\end{align}
Here, $\tilde{D}_\mu$ is the covariant derivative associated with the subgroup $G \subset SU(3)$. The terms $\chi \varsigma$ and $\left( \bar{\varsigma} \varsigma \right) \varsigma$ approximately describe the interaction between the ``purely quantum'' degrees of freedom $A^m_\mu$ and the sea quarks $\widehat{\delta \psi}$. 

\begin{figure}[H]
    \begin{center}
        \includegraphics[width=.49\linewidth]{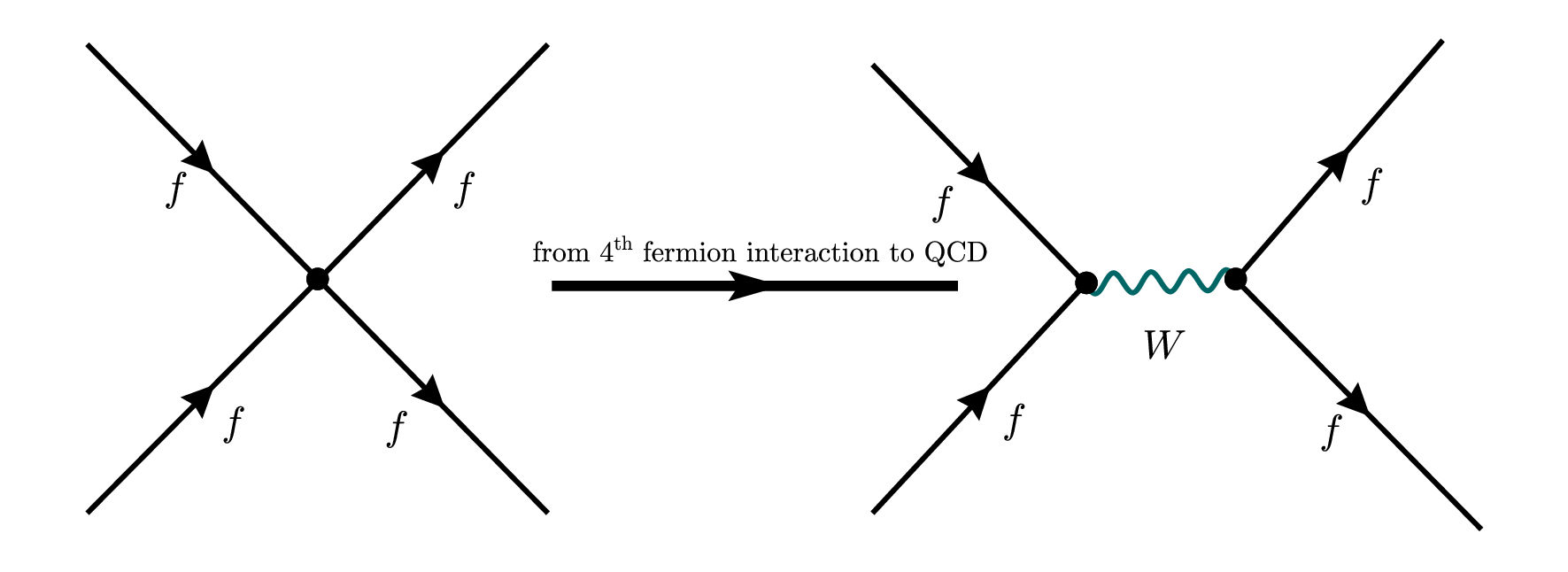}
    \end{center}
    \vspace{-.5cm}
    \caption{Transition from a four-fermion theory to gauge interactions between fermions.}
\label{4_fermions_gauge}
\end{figure}

To understand why the nonlinear term
$
\varsigma \left( \bar{\varsigma}\varsigma \right)
$
appears on the right-hand side of Eq's~\eqref{5_a_45},\eqref{5_a_47} and \eqref{5_B_14}, it is useful to recall that, before the development of gauge-field theories, fermion interactions were described by four-fermion theories, in which the interaction between fermions was represented by a vertex involving four fermion fields; see the left-hand panel of Fig.~\ref{4_fermions_gauge}. Gauge theories describe this interaction in terms of the exchange of a gauge-field quantum; see the right-hand panel of Fig.~\ref{4_fermions_gauge}. Thus, the transition from a four-fermion interaction theory to a gauge-field description can be understood schematically from Fig.~\ref{4_fermions_gauge}.

If the figure is read in the reverse direction, it suggests that an approximate nonperturbative description of the interaction 
$
	\widehat{\overline{\delta \psi}}(y)	A^m_\mu \gamma^\mu \lambda^a
 \widehat{\delta \psi}
$ entering the left-hand side of Eq.~\eqref{5_B_10} can be given by the expression schematically written as 
\begin{equation}
	\expval{ \widehat{\overline{\delta \psi}}(y)
		A^m_\mu \gamma^\mu \lambda^a
		 \widehat{\delta \psi}
	} \approx 
	\left[ \bar{\varsigma}(y) \varsigma \right] \left( \bar{\varsigma}\varsigma \right) . 
\label{5_B_50}
\end{equation}
Thus, the approximate nonperturbative expression \eqref{5_B_10} for the interaction \eqref{5_B_50} leads us to a model of the Nambu--Jona-Lasinio type. \textit{The approximation \eqref{5_B_50} has a clear physical interpretation: it describes the quark condensate $\varsigma$ formed by sea (virtual) quarks bound by the vacuum gauge field $A^m_\mu$.}

The Dirac equation \eqref{5_B_12} describes the interaction of fermions with the ``almost classical'' degrees of freedom of the $G \subset SU(3)$ gauge field, whereas Eq.~\eqref{5_B_14} describes the interaction of sea quarks with the ``almost quantum'' degrees of freedom of the gauge field.

\subsection{Physical assumptions for the Dirac equation and currents}
\label{quark_assumptions}

In deriving the approximations for the Dirac equation and the currents, we have made the following assumptions: 
\begin{itemize}
\item In the vacuum case: 
\begin{itemize}
\item Bilinear combinations of the quark-field triplet $\psi$ are approximately represented by bilinear combinations of the spinor singlet $\varsigma$, which describes the sea-quark condensate.
\item The three-point Green's function $\expval{\bar{\psi}(y) \gamma^\mu A^B_\mu \lambda^B \psi}$ arising in the Dirac equation is approximately represented as the sum of two terms. The first term is the product of the quantity describing the gauge-field condensate $\phi$ and the quantity describing the sea-quark condensate $\varsigma$, whereas the second term is the square of the sea-quark condensate. Schematically, this can be written as follows: 
$
	\expval{\bar{\psi}(y) \gamma^\mu A^B_\mu \lambda^B \psi} \approx 
	\phi \bar{\varsigma}(y) \varsigma_\alpha 
	+ \bar{\varsigma}(y) \varsigma 
	\left( 
		\bar{\varsigma} \varsigma
	\right)
$, see \eqref{5_a_40}. 
\item The three-point Green's function $\expval{\bar{\psi} \gamma^\mu A^B_\mu(y) \lambda^B \psi}$ arising in the treatment of the currents \eqref{YM_scalar_appr_final}, \eqref{3_C_10}, and \eqref{3_C_20} is schematically approximated as
$
	\expval{\bar{\psi} \gamma^\mu A^B_\mu(y) \lambda^B \psi} 
	\approx \phi(y) \bar{\varsigma} \varsigma
$, see \eqref{5_C_30}, \eqref{5_C_32}.
\end{itemize}
\item In the nonvacuum case: 
\begin{itemize}
\item Bilinear combinations of the quark-field triplet $\psi$ are represented as bilinear combinations of the triplet $\vartheta = \expval{\psi}$ and the spinor triplet $\varsigma$, which describes the sea-quark condensate.
\item The three-point Green's function
$
\expval{\bar{\psi}(y) \gamma^\mu A^a_\mu \lambda^a \psi}
$
arising in the treatment of the Dirac equation \eqref{5_B_10} can be schematically approximated as
$
\expval{\bar{\psi}(y) \gamma^\mu A^a_\mu \lambda^a \psi} \approx
\bar{\vartheta}(y) A^a_\mu \gamma^\mu \lambda^a \vartheta
+ \bar{\varsigma}(y) \left( A^a_\mu \gamma^\mu \lambda^a
+ \kappa_1 \chi +
\kappa_2 \bar{\varsigma} \varsigma
\right) \varsigma,
$
i.e., it is decomposed into two parts: the first contains only $\vartheta$, whereas the second contains only $\varsigma$; see \eqref{В_80}. This allows us to split the Dirac equation into two equations: a linear Dirac equation describing the expectation value $\vartheta$, and a nonlinear Dirac equation describing the sea-quark condensate $\varsigma$. 
\item The three-point Green's function
$
\expval{\hat{\bar{\psi}} \gamma^\mu \lambda^a \hat{\psi}},
$
arising in the treatment of the currents in Eq.~\eqref{3_D_100}, can be schematically approximated as
$
 \expval{\hat{\bar{\psi}} \gamma^\mu \lambda^a \hat{\psi}} 
	\approx 
	\bar{\vartheta} \gamma^\mu \lambda^a  \vartheta 
	+ \bar{\varsigma} \gamma^\mu \lambda^a  \varsigma 
$, see \eqref{В_130}. 
\item The three-point Green's function
$
\expval{\bar{\psi} \gamma^\mu A^m_\mu(y) \lambda^m \psi},
$
arising in the treatment of the currents in Eq.~\eqref{3_D_110}, can be schematically approximated as
$
	\expval{\bar{\psi} \gamma^\mu A^m_\mu(y) \lambda^m \psi}
	\approx \chi (y)\bar{\varsigma} \varsigma 
$, see \eqref{В_140}. 
\end{itemize}
\end{itemize}

\section{Final equations for the truncated system of nonperturbative Dyson--Schwinger equations}
\label{final_eqns}

In this section, we present the final equations in the first approximation to the nonperturbative Dyson--Schwinger equations, obtained previously in Secs.~\ref{vacuum_one_field}, \ref{two_fields}, and \ref{quark_approx}.

The system of equations describing the vacuum case, in which $\expval{A^B_\mu} = \expval{\psi} = 0$, $(B = 1, 2, \ldots, 8)$, takes the following form: 
\begin{align}
		\lambda_{\mu \nu} \partial^\mu \partial^\nu \phi - \lambda \Box \phi 
		+ \Lambda \left( \phi^2 - \phi_0 \right) \phi 
	& = - j_0 \bar{\varsigma} \varsigma , 
\label{6_20}\\
	\imath 
	\qty(\gamma^\mu)_{\alpha \beta}
	\partial_\mu \varsigma_{\beta} 
	+ \delta_1 \phi \varsigma_\alpha 
	+ \delta_3 \varsigma_{\alpha}  \left( 
			\bar{\varsigma}_\beta \varsigma_\beta
		\right) 
	- m \varsigma_{\alpha} & = 0 . 
\label{6_22}
\end{align}
Here, the scalar function $\phi$ describes the two-point Green's function
$
\expval{A^B_\mu(y) A^C_\nu},
$
which is approximately represented as the product of the same scalar function $\phi$ evaluated at the points $x$ and $y$; details are given in Sec.~\ref{one_field}. The spinor $\varsigma_{\alpha}$ approximately describes the two-point Green's function $\expval{\hat{\bar{\psi}}_{\alpha i}(y) \hat{\psi}_{\beta i}}$; details are given in Sec.~\ref{vacuum_quark}. The derivation of the Dirac equation for the spinor field $\varsigma$ is presented in Sec.~\ref{Dirac_appr}.

The system of equations describing the vacuum case, in which
$\expval{A^a_\mu} = \expval{A^m_\mu} = 0$, but the two-point Green's functions for $A^a_\mu$ and $A^m_\mu$ are approximately described by different scalar fields $\phi$ and $\chi$, while $\expval{\psi} = 0$, takes the following form:
\begin{align}
	\lambda_{\mu \nu} \partial^\mu \partial^\nu \phi - \lambda \Box \phi 
	+ \left( 
				\lambda_1 \phi^2 + \lambda_2 \chi^2 + \lambda_3
			\right) \phi & = - j_1 \bar{\varsigma} \varsigma , 
\label{6_30}\\
	\xi_{\mu \nu} \partial^\nu \partial^\mu \chi - \xi \Box \chi 
	+ \left( 
			\xi_1 \chi^2 + \xi_2 \phi^2 + \xi_3
	\right) \chi  & = - j_2 \bar{\varsigma} \varsigma , 
\label{6_40}\\
	\imath 
	\qty(\gamma^\mu)_{\alpha \beta}
	\partial_\mu \varsigma_{\beta} 
	+ \left( \delta_1 \phi 	+ \delta_2 \chi \right) \varsigma_\alpha 
	+ \delta_3 \varsigma_{\alpha}  \left( 
			\bar{\varsigma}_\beta \varsigma_\beta
		\right) 
	- m \varsigma_{\alpha}& = 0 . 
\label{6_50}
\end{align}
Here, the scalar fields $\phi$ and $\chi$ approximately describe the two-point Green's functions $\expval{A^b_\mu(y) A^c_\nu}$ and $\expval{A^m_\mu(y) A^n_\nu}$, respectively; details are given in Sec.~\ref{two_condensates}. The physical and mathematical meaning of the spinor $\varsigma$ is the same as in Eq.~\eqref{6_22}.

A few remarks are in order concerning the unexpected terms of the form
$\lambda_{\mu \nu} \partial^\mu \partial^\nu \phi$
appearing in Eqs.~\eqref{6_20}, \eqref{6_30}, and \eqref{6_40}. According to \eqref{app_A_60}, these terms arise from an expression of the form
$
A^B_\mu(y) \partial^\mu \partial_\nu A^{B \nu}(x).
$
In classical field theories, one can impose the gauge condition $\partial_\nu A^{B \nu} = 0$, in which case these terms vanish. Whether such a gauge condition can be imposed in the present case remains unclear. 

The system of equations describing the case with ``almost classical'' and ``almost quantum'' degrees of freedom takes the following form:
\begin{align}
	\widetilde{D}_\nu	\mathcal{F}^{a \mu \nu} 
	+ g \widetilde{D}_\nu \left( \zeta^{a \mu \nu} \chi^2 \right) 
	- \frac{g}{2} \widetilde{D}_\nu \left( \zeta^{a \nu \mu} \chi^2 \right) 
	- \frac{g}{2} \widetilde{D}^\mu \left( \zeta^{a} \chi^2 \right) 
	+ \xi^{a b \mu \nu} A^b_\nu \chi^2 
	& = 
		- \left( 
			\bar{\vartheta} \gamma^\mu\lambda^a \vartheta 
			+ \bar{\varsigma} \gamma^\mu \lambda^a  \varsigma 
		\right) , 
\label{6_60}\\
	\zeta^{\mu \nu} \partial_\mu \partial_\nu \chi 
	+ g \zeta^{a \mu \nu} \mathcal{F}^a_{\mu \nu} \chi 
	+ g \nabla_\nu 
	\left( 
		 \zeta^a A^{a \nu} \chi - \zeta^{a \mu \nu} A^a_\mu \chi  
	\right) &
\nonumber\\
	- g 
	\left( 
		\zeta^{a \mu \nu} A^a_\nu - \zeta^a A^{a \mu} 
		\right) \nabla_\mu \chi 
		+ \Lambda \left( 
			\chi^2 - \chi_0
	\right) \chi 
	- m^{ab \mu \nu} A^a_\nu A^b_\mu \chi 
	& = - \kappa_1 \bar{\varsigma} \varsigma , 
\label{6_70}\\
	\left[
		\imath \qty(\gamma^\mu)_{\alpha \beta}
		\qty(\tilde{D}_\mu)_{ij} - m \delta_{\alpha \beta} \delta_{i j}
	\right] \vartheta_{\beta j} & = 0 , 
\label{6_80}\\
	\left[ 
		\imath \qty(\gamma^\mu)_{\alpha \beta}
		\qty(\tilde{D}_\mu)_{ij} - m \delta_{\alpha \beta} \delta_{i j} + 
		\left( 
		\kappa_1 \chi 
		- \kappa_2 \bar{\varsigma} \varsigma  
		\right) \delta_{\alpha \beta} \delta_{i j} 
	\right] \varsigma_{\beta j} & = 0 . 
\label{6_85}
\end{align}
Here $\tilde{D}_\mu = \partial_\mu + g f^{abc} A^b_\nu$ is the gauge-covariant derivative in the subgroup $G \subset SU(3)$. The scalar function $\chi$ describes the two-point Green's function $\expval{A^m_\mu(y) A^n_\nu}$ in the same manner as above. We have $\hat{\psi} = \varsigma + \widehat{\delta \psi}$, where the spinor $\varsigma$ is a triplet, in contrast to the vacuum cases. Equation \eqref{6_60} describes the ``almost classical'' degrees of freedom of the $G \subset SU(3)$ gauge field $A^a_\mu$, with the expectation value of the quark field $\vartheta$ and the quark condensate $\varsigma$ acting as sources. Equation \eqref{6_70} describes the gluon condensate $\chi$ generated by the ``almost quantum'' degrees of freedom, with the quark condensate $\varsigma$ acting as its source. The Dirac equation \eqref{6_80} is an equation for the expectation value of the quark field $\vartheta$. Equation \eqref{6_85} is a nonlinear Dirac equation describing the sea-quark condensate $\varsigma$, with quantum corrections $\chi$ arising from the interaction between the quark and gluon condensates, as well as the term $\bar{\varsigma} \varsigma$ arising in the three-point Green's function $\expval{\bar{\psi} A^m_\mu \psi}$.

Let us note that in the case of weak color asymmetry, i.e., when the coefficients
$
	\zeta^a = f^{amn} \zeta^{mn \phantom{\nu} \nu}_{\phantom{mn} \nu} 
	= f^{amn} \zeta^{[m, n] \phantom{\nu} \nu}_{\phantom{mn} \nu} / 2 
	\ll \zeta^{mn \alpha \beta}
$ and 
$
	\zeta^{a \mu \nu} = f^{amn} \zeta^{mn \mu \nu} 
	= f^{amn} \zeta^{[m, n] \mu \nu} / 2 
	\ll \zeta^{mn \alpha \beta}
$  (here, $[,]$ denotes antisymmetrization), in Eqs.~\eqref{6_60} and \eqref{6_70}, the terms containing these coefficients can then be neglected, yielding the following simplified form of the equations:
\begin{align}
	\widetilde{D}_\nu	\mathcal{F}^{a \mu \nu} 
	+ \xi^{a b \mu \nu} A^b_\nu \chi^2 
	& = 
		- \left( 
			\bar{\vartheta} \gamma^\mu\lambda^a \vartheta 
			+ \bar{\varsigma} \gamma^\mu \lambda^a  \varsigma 
		\right) , 
\label{6_90}\\
	\zeta^{\mu \nu} \partial_\mu \partial_\nu \chi 
	+ \Lambda \left( 
		\chi^2 - \chi_0
	\right) \chi 
	- m^{ab \mu \nu} A^a_\nu A^b_\mu \chi 
	& =	- \kappa_1 \bar{\varsigma} \varsigma . 
\label{6_100}
\end{align}

The physical meaning of these equations is as follows:
\begin{itemize}
\item Equations \eqref{6_20}, \eqref{6_22}, and \eqref{6_30}--\eqref{6_50} describe the vacuum condensates of gluon fields and quarks (sea quarks).
\item In a nonperturbative vacuum, in contrast to a perturbative vacuum, quantum soliton-like fluctuations may arise. These fluctuations are described by Eqs.~\eqref{6_20}, \eqref{6_22}, or \eqref{6_30}--\eqref{6_50}. There is some analogy here with lattice calculations, which indicate that the dominant contribution to the QCD path integral comes from field configurations containing monopole-like objects (hedgehogs).
\item Equations \eqref{6_60}--\eqref{6_85} describe a physical system in which the quantum $SU(3)$ gauge fields can be approximately decomposed into an ``almost classical'' Yang--Mills--Proca field $\tilde{F}^a_{\mu \nu}$, sourced by real quarks described by the Dirac equation \eqref{6_80}, and an ``almost quantum'' field described by the scalar field $\chi$ (the gluon condensate), sourced by sea quarks described by the nonlinear Dirac equation \eqref{6_85}.
\item Equations \eqref{6_60}--\eqref{6_85} can describe soliton-like quantum fluctuations that additionally contain color electric and magnetic fields. These equations can also describe quantum-averaged color electric and magnetic fields between dynamical quarks inside a hadron.
\end{itemize}

It is worth emphasizing the appearance of a nonlinear Dirac equation in all these equations. In the 1950s, Refs.~\cite{Finkelstein:1951zz, Finkelstein:1956} showed that the nonlinear Dirac equation possesses soliton-like solutions as well as a global minimum in the energy spectrum of its solutions. The authors referred to the corresponding solution as ``the lightest stable particle.'' In modern terminology, this corresponds to a mass gap. Refs.~\cite{Dzhunushaliev:2020qwf, Dzhunushaliev:2021apa} demonstrated that a mass gap and soliton solutions of the same type also exist in Yang--Mills theory coupled to a nonlinear Dirac equation. This indicates that the properties of the nonlinear Dirac equation associated with the existence of soliton-like solutions and a mass gap are preserved when the fermionic field interacts with other fields. This provides strong grounds for expecting that the mass gap will likewise persist in our case.
An intriguing question in this context is whether this mass gap persists at higher orders of approximation and, in particular, what happens in the limiting transition to the infinite system of nonperturbative Dyson--Schwinger equations? If the answer is affirmative, this would indicate that the nonlinear Dirac equation provides a sufficiently accurate approximation to the properties of the interaction between the gauge field and quarks in the infinite system of nonperturbative Dyson--Schwinger equations.

A natural question arising in the analysis of the systems of equations \eqref{6_20}--\eqref{6_22}, \eqref{6_30}--\eqref{6_50}, and \eqref{6_60}--\eqref{6_85} is how the validity of these approximations can be tested? We believe that this can be done by comparing the solutions obtained from these systems with lattice calculations. For example, in Refs.~\cite{Dzhunushaliev:2024dzp, Dzhunushaliev:2025olg}, we compared the distributions of color electric fields and the corresponding energy spectra obtained by solving the strongly simplified Eq.~\eqref{6_60} with ad hoc introduced sources in the form of static quarks with analogous results from lattice calculations. It was found that the results obtained in Refs.~\cite{Dzhunushaliev:2024dzp, Dzhunushaliev:2025olg} are in satisfactory agreement with the results obtained from lattice calculations.

\section{Gauge invariance
}
\label{gauge_inv}

A highly challenging issue in nonperturbative quantum chromodynamics is gauge invariance. The difficulty lies in demonstrating local gauge invariance of the infinite system of Dyson--Schwinger equations. In our case, it appears that truncating this system to a finite set of equations effectively amounts to choosing a gauge from the outset.

In the vacuum case considered in Secs.~\ref{one_field} and \ref{two_condensates}, the gauge is chosen as follows:
\begin{equation}
\expval{A^{B \mu}} = 0 ,
\label{7_10}
\end{equation}
which is consistent with the conditions $\expval{A^{B \mu}} = 0$, see \eqref{app_A_5} and \eqref{app_B_5}.
In this case, the following relation holds:
\begin{equation}
\expval{\nabla_\mu A^{B \mu} (x)} = \expval{\partial_\mu A^{B \mu}(x)} +
g f^{BCD} \expval{A^C_\mu(x) A^{D \mu}(x)} = 0.
\label{7_15}
\end{equation}
The second term in \eqref{7_15} vanishes identically because the structure constants $f^{BCD}$ are completely antisymmetric, whereas the expression $A^C_\mu(x) A^{D \mu}(x)$ is symmetric under the interchange of the indices $C$ and $D$ in the vacuum case.

In the asymmetric case considered in Sec.~\ref{two_fields}, the gauge choice is defined by the following conditions: (a) all degrees of freedom $A^B_\mu$ are divided into ``almost classical'' degrees of freedom $A^a_\mu$, belonging to the subgroup $A^a_\mu \in G \subset SU(3)$, and ``almost quantum'' degrees of freedom $A^m_\mu$, belonging to
$A^m_\mu \in SU(3) / G$; (b) for the ``almost quantum'' degrees of freedom, we choose a gauge analogous to \eqref{7_10}:
\begin{equation}
\expval{A^{m \mu}} = 0 .
\label{7_20}
\end{equation}
Similarly to \eqref{7_15}, we then have
\begin{equation}
\expval{\nabla_\mu A^{m \mu} (x)} = \expval{\partial_\mu A^{m \mu}(x)} .
\label{7_25}
\end{equation}

\section{Turbulence modeling and nonperturbative quantization}
\label{turb_QCD}

In hydrodynamics, the modeling of turbulent fluid motion gives rise to the so-called closure problem \cite{Wilcox}. This problem arises because the modeling requires determining the mean values of velocities and pressures, their dispersions, and so on. To this end, the Navier--Stokes equations are averaged; however, this procedure generates mean values of products of velocities, pressures, and other quantities. To derive equations for these quantities, the Navier--Stokes equations are multiplied by the corresponding variables and then averaged. Unfortunately, these new equations contain additional unknown functions (cumulants), and this process continues indefinitely. As a result, one obtains an infinite system of equations for all cumulants.
Clearly, such a system cannot be solved exactly, and the problem of finding an approximate solution arises. The most natural way to treat such an infinite system of equations is to truncate it to a finite system. This, however, raises the question of how to treat the higher-order cumulants appearing in the last equation? Some approximation for these cumulants must be introduced. For example, one may assume that (1)~all higher-order cumulants are negligibly small compared with lower-order ones, or (2)~the highest-order cumulant retained in the truncated system is represented as a combination of lower-order cumulants. In both cases, one obtains a finite system of equations for all cumulants up to order $n$.

In the nonperturbative quantum field theory considered here, a similar problem arises: there is an analogous infinite system of equations for all Green's functions (see, e.g., Refs.~\cite{heis, Dzhunushaliev:2022apb}). Thus, similarly to the situation encountered in the stochastic modeling of turbulence, a closure problem arises in the nonperturbative approach. This problem concerns how to eliminate certain higher-order Green's functions from the last equations of the truncated system of Dyson--Schwinger equations? Here, we resolve this problem by approximating higher-order Green's functions in terms of lower-order Green's functions.

In this work, we propose possible truncation schemes for certain Green's functions and show that this procedure leads to the appearance of dimensionful constants (closure constants). The emergence of these constants gives rise to the phenomenon of dimensional transmutation in nonperturbative quantum field theory.

\section{Discussion and conclusions}
\label{conclusions}

In this study, we propose a procedure for truncating the infinite system of nonperturbative Dyson--Schwinger equations for quantum chromodynamics. Such a truncation is essential, since there is virtually no prospect of obtaining an exact analytical solution of an infinite system of partial differential equations for Green's functions that, in addition, depend on several variables. We consider this procedure in two cases:
\begin{itemize}
\item In the first case, we consider the vacuum state, in which the quantum-field-theoretic expectation values of the gauge and fermion fields vanish.
\begin{itemize}
\item Therefore, for the gauge field, it is necessary to consider the equation for the two-point Green's function. The main idea is to approximate the two-point Green's function by the product of the same scalar function evaluated at the two corresponding points. To derive an equation for this scalar field, we consider the contraction of the first equation in the Dyson--Schwinger hierarchy. This equation is not closed because it contains a four-point Green's function. To close the equation, we assume that the four-point Green's function can be approximately represented as a bilinear combination of two-point Green's functions. This gives rise to the so-called closure constants.
\item For the currents appearing on the right-hand sides of the Yang--Mills equations, we propose an approximation in which the expectation value of a bilinear combination of the quark-field triplet is approximated by a bilinear combination of a singlet spinor field.
\item The Dirac equation for the triplet of quark fields is then approximated by a Dirac equation for this singlet spinor field.
\end{itemize}

\item In the second case, we consider a physical system with color/spacetime asymmetry. Such an asymmetry must arise when external sources for the gauge field, namely quarks, are present. In this case, color fields arise between the quarks: a longitudinal color electric field in a flux tube connecting a quark and an antiquark, or a $Y$-shaped color electric field between three quarks in a hadron. We assume that such a quantum-field-theoretic system contains ``almost classical'' degrees of freedom describing the aforementioned color electric fields and belonging to some subgroup $G \subset SU(3)$, as well as ``almost quantum'' degrees of freedom belonging to $SU(3) / G$.
\begin{itemize}
\item For the ``almost classical'' degrees of freedom, we obtain a classical equation with quantum corrections, which takes the form of a Proca-like equation. \item The sources (currents) for the ``almost classical'' degrees of freedom are provided both by the expectation value of the quark field and by the quantum condensate of sea quarks.
\item For the ``almost quantum'' degrees of freedom, we obtain an equation for a scalar field that approximately describes the two-point Green's function of these degrees of freedom. \item The source (current) for the ``almost quantum'' degrees of freedom is the quantum condensate of sea quarks.
\item In this case, the original Dirac equation is split into a linear Dirac equation for the expectation value of the spinor field describing the quarks and a nonlinear Dirac equation describing the sea quarks.
\end{itemize}

\item In both cases, the purely quantum nonlinear term in the Dirac equation describes the interaction between the sea quarks and the vacuum gauge field.
\end{itemize}

Within this approach to solving the system of nonperturbative Dyson--Schwinger equations, namely, by truncating the infinite system to a finite one, the so-called closure constants inevitably appear. They arise when higher-order Green's functions are approximated by some multilinear combinations of lower-order Green's functions. Naturally, these constants are dimensionful, leading to a very interesting situation in which dimensionful constants that are not present in the original Lagrangian emerge upon quantization.
The central question in this context is whether these constants survive in the limiting transition to the infinite system of equations. If the answer is affirmative, i.e., if they do survive, we obtain the phenomenon of ``dimensional transmutation,'' namely, the emergence of new dimensionful constants that are absent from the original Lagrangian; an example is $\Lambda_{\text{QCD}}$ in QCD.
From a mathematical point of view, this means that an infinite system of differential equations is fundamentally different from a finite system of equations. At a qualitative level, the emergence of ``dimensional transmutation'' can be understood as follows. Suppose that we have obtained a solution within a truncated system of equations. Then, for the procedure to converge as higher-order approximations are considered, the constants introduced at the previous approximation level must be preserved in one form or another.

It should be emphasized that the approximation proposed here is applicable only to a static physical system in which all observables are time-independent. For a dynamical system, the situation becomes considerably more complicated, since the finite propagation speed of the fields must be taken into account. In this case, the Green's functions should vanish outside the light cone, which inevitably leads to the appearance of Dirac $\delta$-functions and their derivatives, i.e., to the emergence of distributions, as occurs in perturbative quantum field theory, where only propagating quanta are present.

In perturbative quantum field theory, the presence of Planck's constant $\hbar$ is an essential feature of any quantum formula. As can be seen from the approximation to the infinite system of quantum Dyson--Schwinger equations presented here, the appearance of ``closure constants'' can likewise serve as an indication that a given formula is genuinely quantum. If these constants survive in the transition from the finite truncated system to the infinite one, this would imply that, in nonperturbative quantum field theory, in addition to Planck's constant $\hbar$, the emergence of new dimensionful constants absent from the original Lagrangian would serve as an indicator of the quantum nature of the system. This is precisely what occurs in QCD, where the scale $\Lambda_{\text{QCD}}$ emerges.

Another interesting feature of this approximation is the emergence of a nonlinear Dirac equation as an equation that approximately describes the three-point Green's function $\bar{\psi} A^B_\mu \lambda^B \gamma^\mu \psi$. It is known that the nonlinear Dirac equation, both as a standalone theory (see Refs.~\cite{Finkelstein:1951zz, Finkelstein:1956}) and when coupled to a Yang--Mills gauge field (see Refs.~\cite{Dzhunushaliev:2020qwf, Dzhunushaliev:2021apa}), possesses a mass gap. This suggests that this property may also persist in our case: the energy spectrum of solutions of any of the systems \eqref{6_20}--\eqref{6_85} may possess a global minimum, corresponding to a mass gap.
The key question is therefore what happens in the limiting transition to the infinite system of Dyson--Schwinger equations: does the mass gap persist? If so, this would indicate that the nonlinear Dirac equation captures the essential properties of the infinite system of Dyson--Schwinger equations!

We consider it very important that the systems of equations \eqref{6_20}--\eqref{6_85} contain a nonlinear Dirac equation. As noted above (see Refs.~\cite{Finkelstein:1951zz}--\cite{Dzhunushaliev:2021apa}), this equation possesses regular soliton-like solutions with finite energy. It is quite plausible that this property is also preserved for Eqs.~\eqref{6_20}--\eqref{6_85}, i.e., that these equations admit regular finite-energy solutions describing configurations of color electric and magnetic fields, for example, between two quarks in a flux tube or a $Y$-shaped configuration between three quarks. This would indicate that the sea-quark condensate plays a central role in the formation of such configurations.

We also note the following circumstance. In nonperturbative quantization, an infinite system of Dyson--Schwinger equations arises for all Green's functions, much as in turbulence modeling (see Ref.~\cite{Wilcox}), where an infinite system of equations is obtained for all cumulants describing correlations of velocities, pressure, and other quantities, leading to similar problems, such as the closure problem. It is therefore natural to ask whether there is some deeper connection between the quantization of strongly nonlinear fields and the stochastic approach to turbulence theory? 

In connection with this issue, we note that it was recently announced \cite{Naview_Stokes} that AI has found a solution to the Navier--Stokes equations that develops a finite-time blowup. This would constitute a counterexample to the global existence and smoothness problem for the Navier--Stokes equations.
This result may provide evidence for a possible connection between turbulence modeling and nonperturbative quantization in the following sense. The solution found in \cite{Naview_Stokes} becomes singular in finite time, i.e., the fluid velocities become unbounded. Apparently, this implies that, as the system approaches the time at which the velocities diverge, the flow should transition to a turbulent regime. However, an adequate description of the turbulent regime requires turbulence modeling based on averaging the Navier--Stokes equations and introducing an infinite system of equations for all cumulants.
Thus, the description of laminar fluid motion at Reynolds numbers below the critical value, $\text{R} < \text{R}_{\text{cr}}$, should be based on the Navier--Stokes equations, whereas the description of turbulent fluid motion at $\text{R} > \text{R}_{\text{cr}}$ requires a turbulence model based on an infinite system of equations for all cumulants. This situation is analogous to the transition from classical to quantum theory. In the classical limit, $\hbar = 0$, one has a classical theory, which may be associated with laminar flow at $\text{R} < \text{R}_{\text{cr}}$, whereas for $\hbar \neq 0$ one has a quantum theory, which may be associated with turbulent flow at $\text{R} > \text{R}_{\text{cr}}$. This means that Planck's constant $\hbar$ and the critical Reynolds number $\text{R}_{\text{cr}}$ play analogous roles: they separate the quantum regime from the classical one in the first case, and turbulent motion from laminar flow in the second.

 We expect that the finite approximation to the infinite system of nonperturbative Dyson--Schwinger equations obtained here can be applied to: (a) describing color electric and magnetic fields between two quarks in a flux tube, as well as $Y$-shaped configurations of these fields between three quarks; (b) constructing a model of a glueball composed of gauge fields and sea quarks; (c) modeling soliton-like quantum fluctuations in the QCD vacuum; (d) modeling hedgehog configurations in QCD; and so on.
 
 In conclusion, we emphasize that the nonperturbative quantization à la Heisenberg considered here remains largely terra incognita, involving numerous open problems and being, from a mathematical point of view, considerably more complicated than perturbative quantization based on Feynman diagrams. This situation is closely analogous to the distinction between nonlinear and linear classical theories. For linear theories, one can generally write down a general solution, whereas for nonlinear theories this is not possible, and a specific approach must be developed for each individual case. The reward, however, is the possibility of describing remarkable objects such as the ’t~Hooft--Polyakov monopole, instantons, solitons, black holes, wormholes, and so on.
Similarly, in the case of nonperturbative quantization à la Heisenberg, one may hope that this approach will provide a correct description of nonperturbative effects in strongly nonlinear fields. In particular, in QCD, this may include confinement, dimensional transmutation, and related phenomena. All of this indicates that the complexity of nonperturbative quantization is considerably greater than that of perturbative quantization.

\section*{Acknowledgements}

We gratefully acknowledge support provided by the program No.~AP26195069 (Bound states in Maxwell, Yang-Mills, Proca theories with and without gravity in the presence of spinor fields) of the Committee of Science of the Ministry of Science and Higher Education of the Republic of Kazakhstan. 

\appendix

\section{Vacuum approximation for $\left\langle A^B_\mu \right\rangle \approx 0$ by a single scalar field
}
\label{one_scalar_field}

In this section we consider the case
\begin{equation}
	\expval{ A^B_\mu} = 0 . 
\label{app_A_5}
\end{equation}
Let us consider the contraction of the Yang--Mills operator equation \eqref{NP_20}
\begin{equation}
	\left\langle A_\mu^B \left( y \right) 
		 F^{B \mu \nu}_{; \nu} \left( x \right) 
	\right\rangle = -  \left\langle 
		\bar {\psi} \left( x \right) A_\nu^B\left( y \right) \lambda^B \gamma^\nu \psi \left( x \right) 
	\right\rangle . 
\label{app_A_10}
\end{equation}
Here and throughout the following, the derivative $\partial_\mu$ acts on the coordinate $x$. To derive the equation for the scalar function $\phi$, we will use the following approximations: 
\begin{align}
	\expval{A^B_\mu(y) A^C_\nu(x)} & \approx 
	\zeta^{B C}_{\phantom{B C}\mu \nu} \phi(y) \phi(x) , 
\label{app_A_20}\\
	\left\langle A^B_\mu(y) A^C_\nu(x) A^D_\rho(x)\right\rangle & \approx 0, 
\label{app_A_30}\\
	\left\langle 
		A^B_\mu(y) A^C_\nu(x) A^D_\rho(x) A^E_\sigma(x) 
	\right\rangle & \approx 
	\left\langle 
		A^B_\mu(y) A^C_\nu(x) 
	\right\rangle
	\left\langle 
		A^D_\rho(x) A^E_\sigma(x) 
	\right\rangle 
	+	\lambda^{D E}_{\phantom{D E}\rho \sigma} 
	\left\langle 
		A^B_\mu(y) A^C_\nu(x) 
	\right\rangle 
\nonumber \\
	& + \left\langle 
		A^B_\mu(y) A^D_\rho(x) 
	\right\rangle
	\left\langle 
		A^C_\nu(x) A^E_\sigma(x) 
	\right\rangle 
	+	\lambda^{C E}_{\phantom{C E}\nu \sigma} 
	\left\langle 
		A^B_\mu(y) A^D_\rho(x) 
	\right\rangle 
\nonumber \\
	& + \left\langle 
		A^B_\mu(y) A^E_\sigma(x) 
	\right\rangle
	\left\langle 
		A^C_\nu(x) A^D_\rho(x) 
	\right\rangle 
	+	\lambda^{C D}_{\phantom{D E}\nu \rho} 
		\left\langle 
			A^B_\mu(y) A^E_\sigma(x) 
	\right\rangle , 
\label{app_A_40}\\
	\left\langle 
		A^B_\mu(y) \partial_\nu A^C_\rho(x)
	\right\rangle & \approx \zeta^{B C}_{\phantom{B C}\mu \rho} \phi(y) \partial_\nu \phi(x) .
\label{app_A_50}
\end{align}
Here, the constants $\zeta^{B C}_{\phantom{B C}\mu \nu}$ are introduced. They are symmetric within each pair of color indices $B,C$ and spacetime indices $\mu,\nu$, respectively, because there is no preferred direction in the vacuum. We also introduce the closure constants $\lambda^{BC}_{\phantom{AB}\mu\nu}$, whose physical meaning  is that they characterize the condensate $\left\langle A^B_\mu(x) A^C_\nu(x) \right\rangle$.
In the static case, in the four-point Green's function
$\expval{A^B_\mu(y) A^C_\nu(x) A^D_\rho(x) A^E_\sigma(x)}$,
the operators $A^C_\nu(x)$, $A^D_\rho(x)$, and $A^E_\sigma(x)$ commute. Therefore, the closure constants
$
\lambda^{D E}_{\phantom{d e}\rho \sigma},
\lambda^{C E}_{\phantom{d e}\nu \sigma},
\lambda^{C E}_{\phantom{d e}\nu \sigma}
$
are symmetric within each pair of color and spacetime indices.

Let us note that the symmetry properties of the coefficients $c^{B C}_{\phantom{B C}\mu \nu}$ can be compared with the situation for velocity correlation functions in the theory of turbulence:
\begin{equation}
B_{i k} = \overline{
\left( v_{2 i} - v_{1 i}\right) \left( v_{2 k} - v_{1 k}\right)
},
\label{app_A_52}
\end{equation}
where $\vec{v}_2$ and $\vec{v}_1$ are the fluid velocities at two neighboring points $\vec{r}_2$ and $\vec{r}_1$, respectively, and the overbar denotes a time average. As stated in the textbook \cite{Landau}, the most general form of such a rank-two tensor is
\begin{equation}
B_{i k} =
A(r) \delta_{i k} + B(r) n_i n_k ,
\label{app_A_55}
\end{equation}
where $\vec{n}$ is the unit vector in the direction of $\vec{r}_2 - \vec{r}_1$. It follows directly from \eqref{app_A_55} that the tensor $B_{i k}$ is not symmetric.
The main difference between $B_{i k}$ and $\left\langle A^B_\mu(x) A^C_\nu(x) \right\rangle$ is that the averaging in
$
\overline{
\left( v_{2 i} - v_{1 i}\right) \left( v_{2 k} - v_{1 k}\right)
}
$
is stochastic, whereas the averaging in $\left\langle A^B_\mu(x) A^C_\nu(x) \right\rangle$ is quantum. Nevertheless, their physical content remains the same: both expressions describe correlations of physical quantities at two points. This observation provides an argument in favor of the statement that the Green's function $\left\langle A^B_\mu(x) A^C_\nu(x) \right\rangle$, similarly to the cumulant $B_{i k}$ in turbulence, is not symmetric with respect to the color and spacetime indices.

The approximation \eqref{app_A_30} implements our assumption that all odd-order Green's functions vanish. The approximation \eqref{app_A_40} is necessary for truncating the infinite system of Dyson--Schwinger equations by approximating higher-order Green's functions in terms of lower-order Green's functions, namely,
$G_4 \approx G_2^2 + \lambda_1 G_2 + \lambda_2$.

On the left-hand side of Eq.~\eqref{app_A_10}, we have the following expression:
\begin{equation}
	\left\langle 
		A^B_\mu(y) \left[ \partial_\nu \left( 
			\partial^\mu A^{B \nu} - \partial^\nu A^{B \mu} 
			+ g f^{BCD} A^{C \mu} A^{D\nu}
		\right) 
	+ g f^{BCD} A^C_\nu 
	\left( 
		\partial^\mu A^{D \nu} - \partial^\nu A^{D \mu} 
		+ g f^{DMN} A^{M \mu} A^{N\nu}
	\right) 
	\right] 
	\right\rangle . 
\label{app_A_60}
\end{equation}
All operators in the square brackets are functions of $x$. To evaluate the expression containing the second derivatives, we use the approximation \eqref{app_A_20}:
\begin{equation}
	\left\langle 
		A^B_\mu(y) \partial_\nu \left( 
			\partial^\mu A^{B \nu} - \partial^\nu A^{B \mu} 
		\right) 
	\right\rangle \approx 
	\phi(y) \left( 
		\zeta^{BB \mu \nu} \partial_\mu \partial_\nu \phi  - 
		\zeta^{BB \phantom{\mu} \mu}_{\phantom{BB} \mu} \Box \phi  
	\right) .
\label{app_A_70}
\end{equation}
According to \eqref{app_A_30}, the quantum expectation value of the product of three operators vanishes:
\begin{equation}
	\left\langle A^B_\mu (y) \partial_\nu 
		\left( A^{C \mu} A^{D\nu} \right) 
	\right\rangle \approx 0, \quad 
	\left\langle 
		A^B_\mu (y) A^C_\nu \left( 
			\partial^\mu A^{D \nu} - \partial^\nu A^{D \mu}
		\right) 
	\right\rangle \approx 0 . 
\label{app_A_80}
\end{equation}
The last term in \eqref{app_A_60} is determined according to the approximation \eqref{app_A_40}: 
\begin{equation}
\label{app_A_90}
\begin{split}
	& g^2 f^{B C D} f^{D M N} \left\langle 
		A^B_\mu(y) A^C_\nu(x) A^{M \mu}(x) A^{N \nu}(x) 
	\right\rangle 
\\
	& \approx g^2 f^{B C D} f^{D M N} 
	\left( 
		\zeta^{B C}_{\phantom{B C}\mu \nu} \zeta^{M N \mu \nu} + 
		\zeta^{B M \phantom{\mu} \mu}_{\phantom{B M}\mu} \zeta^{C N \phantom{\nu} \nu}_{\phantom{CN} \nu} 
		 + \zeta^{B N}_{\phantom{B C}\mu \nu} \zeta^{C M \nu \mu} 
	\right) 
	\phi(y) \phi^3(x) 
\\
	& + g^2 f^{B C D} f^{D M N} 
	\left( 
		\zeta^{B C}_{\phantom{B C}\mu \nu} \lambda^{M N \mu \nu} + 
				\zeta^{B M \phantom{\mu} \mu}_{\phantom{B M}\mu} 
				\lambda^{C N \phantom{\nu} \nu}_{\phantom{CN} \nu} 
				 + \zeta^{B N}_{\phantom{B C}\mu \nu} \lambda^{C M \nu \mu} 
	\right) \phi(y) \phi(x) 
	= \left[ \Lambda \phi^2(x) - \lambda \right] \phi(x) \phi(y) , 
\end{split}
\end{equation}
where the closure constants $\Lambda, \lambda$ are determined as follows:
\begin{align}
	\Lambda & = g^2 f^{B C D} f^{D M N} 
	\left( 
		\zeta^{B C}_{\phantom{B C}\mu \nu} c^{M N \mu \nu} + 
		\zeta^{B M \phantom{\mu} \mu}_{\phantom{B M}\mu} \zeta^{C N \phantom{\nu} \nu}_{\phantom{CN} \nu} 
	 + \zeta^{B N}_{\phantom{B C}\mu \nu} \zeta^{C M \nu \mu} 
	\right), 
\label{app_A_100}\\
	\lambda & = - g^2 
	f^{B C D} f^{D M N} 
	\left( 
		\zeta^{B C}_{\phantom{B C}\mu \nu} \lambda^{M N \mu \nu} + 
		\zeta^{B M \phantom{\mu} \mu}_{\phantom{B M}\mu} 
		\lambda^{C N \phantom{\nu} \nu}_{\phantom{CN} \nu} 
		+ \zeta^{B N}_{\phantom{B C}\mu \nu} \lambda^{C M \mu \nu} 
	\right) . 
\label{app_A_110}
\end{align}
It should be noted that, owing to the antisymmetry of the structure constants and the symmetry of $\zeta^{B C}_{\phantom{B C}\mu \nu} \lambda^{M N \mu \nu}$ and $\lambda^{M N \mu \nu}$, the first terms on the right-hand sides of \eqref{app_A_100} and \eqref{app_A_110} vanish identically. This allows us to write the contraction \eqref{app_A_10} in the following form:
\begin{equation}
	\left[ 
		\lambda_{\mu \nu} \partial^\mu \partial^\nu \phi - \lambda \Box \phi 
		+ \Lambda \left( \phi^2 - \phi_0 \right) \phi 
	\right] \phi(y)= - j_0  
	\left\langle 
		\bar{\psi}\left( x \right) A^B_\mu(y) \gamma^\mu \lambda^B \psi\left( x \right) 
	\right\rangle .
\label{app_A_120}
\end{equation}
Here, using the commutativity of partial derivatives, we have introduced the constant
$
	\lambda_{\mu \nu} = \left( \zeta^{B B}_{\phantom{B B} \mu \nu} 
	+ \zeta^{B B}_{\phantom{B B} \nu \mu} \right)/2
$, $\lambda = \zeta^{BB \phantom{\mu} \mu}_{\phantom{BB} \mu}$ and $\phi_0 = \phi_2 / \Lambda$. 

\section{Vacuum approximation for $\left\langle A^B_\mu \right\rangle \approx 0$  with two scalar fields
}
\label{two_scalar_fields}

In this section we consider the case
\begin{equation}
	\expval{A^a_\mu} = \expval{A^m_\mu} = 0 
\label{app_B_5}
\end{equation}
with $A^a_\mu \in G \subset SU(3)$ and $A^m_\mu \in SU(3)/G$. 

Let us consider the contraction of the equations 
\begin{align}
	\left\langle A_\mu^a \left( y \right) 
		 F^{a \mu \nu}_{; \nu} \left( x \right)
	\right\rangle = - \left\langle 
		\bar{\psi}\left( x \right) A_\mu^a \left( y \right) \lambda^a \gamma^\mu \psi\left( x \right) 
	\right\rangle , 
\label{app_B_10}\\
	\left\langle A_\mu^m \left( y \right) 
		 F^{m \mu \nu}_{; \nu} \left( x \right)
	\right\rangle = -  \left\langle 
		\bar{\psi}\left( x \right) A_\mu^m \left( y \right) \lambda^m \gamma^\mu \psi\left( x \right) 
	\right\rangle , 
\label{app_B_20}
\end{align}
where the following notations have been introduced: 
\begin{eqnarray}
	F^a_{\mu \nu} = 
		\mathcal{F}^a_{\mu \nu} + g f^{a m n} A^m_\mu A^n_\nu 
	=  \mathcal{F}^a_{\mu \nu} + \mathfrak{F}^a_{\mu \nu} , 
\label{app_B_24}\\
	F^m_{\mu \nu} = 
		\partial_\mu A^m_\nu - \partial_\nu A^m_\mu 
		+ g f^{m n a} \left( A^n_\mu A^a_\nu - A^a_\mu A^n_\nu\right) ,
\label{app_B_26}
\end{eqnarray}
where 
$
	\mathcal{F}^a_{\mu \nu} = 
	\partial_\mu A^a_\nu - \partial_\nu A^a_\mu +
	g f^{a b c} A^b_\mu A^c_\nu 
$ is the strength tensor for $A^a_\mu \in G$, 
$
	\mathfrak{F}^a_{\mu \nu} = g f^{a m n} A^m_\mu A^n_\nu 
$, and $A^m_\mu \in SU(3) / G$.

Similarly to Sec.~\ref{one_field}, we introduce the following approximations for the potential 
$A^{a}_\mu \in G \subset SU(3)$:
\begin{align}
	\left\langle A^b_\mu(y) A^c_\nu(x) \right\rangle & \approx 
	\zeta^{b c}_{\phantom{b c}\mu \nu} \phi(y) \phi(x) , 
\label{app_B_30}\\
	\left\langle 
		A^b_\mu(y) A^c_\nu(x) A^d_\rho(x) A^e_\sigma(x) 
	\right\rangle & \approx 
	\left\langle 
		A^b_\mu(y) A^c_\nu(x) 
	\right\rangle
	\left\langle 
		A^d_\rho(x) A^e_\sigma(x) 
	\right\rangle +	\lambda^{d e}_{\phantom{d e}\rho \sigma} 
	\left\langle 
		A^b_\mu(y) A^c_\nu(x) 
	\right\rangle 
\nonumber \\
	& 
	+ \left\langle 
		A^b_\mu(y) A^d_\rho(x) 
	\right\rangle
	\left\langle 
			A^c_\nu(x) A^e_\sigma(x) 
	\right\rangle + \lambda^{c e}_{\phantom{d e}\nu \sigma} 
		\left\langle 
			A^b_\mu(y) A^d_\rho(x)  
		\right\rangle 
	\nonumber \\
		& 
		+ \left\langle 
			A^b_\mu(y) A^e_\sigma(x) 
		\right\rangle
		\left\langle 
				A^c_\nu(x) A^d_\rho(x) 
		\right\rangle + \lambda^{c d}_{\phantom{d e}\nu \rho} 
			\left\langle 
				A^b_\mu(y) A^e_\sigma(x)  
			\right\rangle ,
\label{app_B_50}\\
	\left\langle 
		A^b_\mu(y) \partial_\nu A^c_\rho(x)
	\right\rangle & \approx \zeta^{b c}_{\phantom{b c}\mu \rho} \phi(y) \partial_\nu \phi(x) . 
\label{app_B_60}
\end{align}
And analogous approximations for the potential $A^{m}_\mu \subset SU(3) / G$:
\begin{align}
	\left\langle A^m_\mu(y) A^n_\nu(x) \right\rangle & \approx 
	\zeta^{m n}_{\phantom{m n}\mu \nu} \chi(y) \chi(x) , 
\label{app_B_70}\\
	\left\langle 
		A^m_\mu(y) A^n_\nu(x) A^p_\rho(x) A^q_\sigma(x) 
	\right\rangle & \approx 
	\left\langle 
		A^m_\mu(y) A^n_\nu(x) 
	\right\rangle
	\left\langle 
		A^p_\rho(x) A^q_\sigma(x) 
	\right\rangle +	\lambda^{p q}_{\phantom{p q}\rho \sigma} 
	\left\langle 
		A^m_\mu(y) A^n_\nu(x) 
	\right\rangle 
\nonumber \\
	& + \left\langle 
		A^m_\mu(y) A^p_\rho(x) 
	\right\rangle
	\left\langle 
		A^n_\nu(x) A^q_\sigma(x) 
	\right\rangle +	\lambda^{n q}_{\phantom{p q} \nu \sigma} 
	\left\langle 
		A^m_\mu(y) A^p_\rho(x) 
	\right\rangle 
\nonumber \\
	& + \left\langle 
		A^m_\mu(y) A^q_\sigma(x) 
	\right\rangle
	\left\langle 
		A^n_\nu(x) A^p_\rho(x) 
	\right\rangle +	\lambda^{n p}_{\phantom{p q} \nu \rho} 
	\left\langle 
		A^m_\mu(y) A^q_\sigma(x) 
	\right\rangle , 
\label{app_B_90}\\
	\left\langle 
		A^m_\mu(y) \partial_\nu A^n_\rho(x)
	\right\rangle & \approx \zeta^{m n}_{\phantom{m n}\mu \rho} \chi(y) \partial_\nu \chi(x) . 
\label{app_B_100}
\end{align}
Similarly to \eqref{app_A_30}, the vacuum expectation values of products of three operators: 
\begin{equation}
	\left\langle A^B_\mu(y) A^C_\nu(x) A^D_\rho(x)\right\rangle \approx 0. 
\label{app_B_105}
\end{equation}
The parameters $\zeta^{b c}_{\phantom{b c}\mu \nu}$ and $\zeta^{m n}_{\phantom{b c}\mu \nu}$, similarly to the previous Appendix~\ref{one_scalar_field}, are symmetric with respect to the color and spacetime indices. The physical meaning of the closure constants $\lambda^{bc}_{\phantom{bc}\mu\nu}$ and $\lambda^{mn}_{\phantom{AB}\mu\nu}$ is that they characterize the condensates $\left\langle A^b_\mu(x) A^c_\nu(x) \right\rangle$ and $\left\langle A^m_\mu(x) A^n_\nu(x) \right\rangle$, respectively.
In the static case, in the four-point Green's function $\expval{A^b_\mu(y) A^c_\nu(x) A^d_\rho(x) A^e_\sigma(x)}$, the operators $A^c_\nu(x)$, $A^d_\rho(x)$, and $A^e_\sigma(x)$ commute. Therefore, the closure constants
$
\lambda^{d e}_{\phantom{d e}\rho \sigma},
\lambda^{c e}_{\phantom{d e}\nu \sigma},
\lambda^{c e}_{\phantom{d e}\nu \sigma}
$
are symmetric within each pair of color and spacetime indices. The same applies to the four-point Green's function
$
	\expval{A^m_\mu(y) A^n_\nu(x) A^p_\rho(x) A^q_\sigma(x)}
$. 

We assume that the fields $A^b_\mu$ and $A^m_\nu$ are uncorrelated with each other, and therefore 
\begin{equation}
	\expval{A^b_\mu A^c_\nu A^m_\rho A^n_\sigma } \approx 
	\expval{A^b_\mu A^c_\nu} \expval{A^m_\rho A^n_\sigma }. 
\label{corrn}
\end{equation}
The left-hand side of the contraction  \eqref{app_B_10} has the following form:
\begin{equation}
\begin{split}
	& \left\langle 
		A^a_\mu(y) \left\lbrace \partial_\nu \left( 
				\partial^\mu A^{a \nu} - \partial^\nu A^{a \mu} +
				g f^{a b c} A^{b \mu} A^{c \nu} + g f^{a m n} A^{m \mu} A^{n \nu} 
			\right) 
	\right. \right. 
\\
	& \left. \left. 
	+ g \left[  
		f^{abc} A^b_\nu \left( \mathcal{F}^{c \mu \nu} + \mathfrak{F}^{c \mu \nu}\right) 
		+ g f^{amn} A^m_\nu F^{n \mu \nu}
	\right] 
	\right\rbrace   
	\right\rangle . 
\label{app_B_120}
\end{split}
\end{equation}
According to the approximation \eqref{app_B_30}, the second derivatives in \eqref{app_B_120} take the following form: 
\begin{equation}
	\expval{
		A^a_\mu(y)  \partial_\nu \left( 
			\partial^\mu A^{a \nu} - \partial^\nu A^{a \mu}
		\right) 
	} \approx \phi(y) \left[ 
		\zeta^{aa}_{\phantom{aa} \mu \nu} \partial^\nu \partial^\mu \phi 
		- \zeta^{aa \phantom{\mu} \mu}_{\phantom{aa} \mu}
	\Box \phi 
	\right] . 
\label{app_B_130}
\end{equation}
According to the approximation \eqref{app_B_105}, the quantum expectation value of any odd-order product of the fields $A^B_\mu$ vanishes
\begin{equation}
\begin{split}
	\expval{A^a_\mu(y) A^{b \mu} A^{c \nu}} \approx 0, 
	\expval{A^a_\mu(y) A^{m \mu} A^{n \nu} } & \approx  0 , 
\\
	\expval{A^a_\mu(y) A^b_\nu
	\left( \partial_\mu A^a_\nu - \partial_\nu A^a_\mu \right) 
	} \approx 0 , 
	\expval{A^a_\mu(y) A^m_\mu \left( \partial^\mu A^{m \nu} - \partial^\nu A^{m \mu} \right)} 
	& \approx 0 . 
\end{split}
\label{app_B_140}
\end{equation}
The following nonzero terms remain:
\begin{align}
	& g^2 f^{abc} f^{cde} \expval{A^a_\mu(y) A^b_\nu A^{d \mu} A^{e \nu}} \approx 
	g^2 f^{abc} f^{cde} \left( 
		\zeta^{ab}_{\phantom{ab} \mu \nu} \zeta^{de \mu \nu} + 
		\zeta^{ad \phantom{\mu} \mu}_{\phantom{ad} \mu} \zeta^{be \phantom{\nu} \nu}_{\phantom{be} \nu} +
		\zeta^{ae \mu \nu} \zeta^{bd}_{\phantom{ae} \nu \mu}
	\right) \phi(y) \phi^3  
\nonumber \\
	& + g^2 f^{abc} f^{cde} \left( 
		\zeta^{ab}_{\phantom{ab} \mu \nu} \lambda^{de \mu \nu} + 
		\zeta^{ad \phantom{\mu} \mu}_{\phantom{ad} \mu} \lambda^{be \phantom{\nu} \nu}_{\phantom{be} \nu} +
		\zeta^{ae \mu \nu} \lambda^{bd}_{\phantom{ae} \nu \mu}
	\right) \phi (y) \phi , 
\label{app_B_145}\\
	& g f^{abc} \expval{A^a_\mu(y) A^b_\nu \mathfrak{F}^{c \mu \nu}} = 
	g^2 f^{abc} f^{c m n} \expval{A^a_\mu(y) A^b_\nu A^{m \mu} A^{n \nu}} 
	\approx g^2 f^{abc} f^{c m n} \zeta^{a b}_{\phantom{a b} \mu \nu} \zeta^{m n \mu \nu} 
	\phi(y) \phi(x) \chi^2(x) , 
\label{app_B_150}\\
	& g^2 f^{amn} f^{npb} \expval{
		A^a_\mu(y) A^m_\nu \left( 
			A^{p \mu} A^{b \nu} - A^{b \mu} A^{p \nu}
		\right) 
	} \approx g^2 f^{amn} f^{npb} \left( 
		\zeta^{ab \phantom{\mu} \nu}_{\phantom{ab} \mu} \zeta^{mp \phantom{\nu} \mu}_{\phantom{mp} \nu} 
		- \zeta^{ab \phantom{\mu} \mu}_{\phantom{ab} \mu} \zeta^{mp \phantom{\nu} \nu}_{\phantom{mp} \nu}
	\right) \phi(y) \phi(x) \chi^2(x) . 
\label{app_B_160}
\end{align}
Owing to the antisymmetry of the structure constants $f^{abc}$ and the symmetry of the closure constants $\zeta^{ab \mu \nu}$ and $\lambda_{ab \mu \nu}$, the first terms on the right-hand sides of \eqref{app_B_145}, \eqref{app_B_150}, and \eqref{app_B_160} vanish.

All of this reduces the contracted equation \eqref{app_B_10} to the following form:
\begin{equation}
	\left[ 
	\lambda_{\mu \nu} \partial^\mu \partial^\nu \phi - \lambda \Box \phi 
	+ \left( 
		\lambda_1 \phi^2 + \lambda_2 \chi^2 - \phi_0
	\right) \phi \right] \phi(y) = 
	- \left\langle 
		\bar{\psi}\left( x \right) A_\mu^a \left( y \right) \lambda^a \gamma^\mu \psi\left( x \right) 
	\right\rangle ,
\label{app_B_170}
\end{equation}
where 
\begin{align}
	\lambda_{\mu \nu} & = \lambda_{\nu \mu}  = 
	\frac{\zeta^{aa}_{\phantom{aa} \mu \nu} + \zeta^{aa}_{\phantom{aa} \nu \mu}}{2} , 
	\quad \lambda = \zeta^{aa \phantom{\mu} \mu}_{\phantom{aa} \mu} , 
\label{app_B_175}\\
	\lambda_1 & = g^2 f^{abc} f^{cde} \left( 
	\zeta^{ad \phantom{\mu} \mu}_{\phantom{ad} \mu} \zeta^{be \phantom{\nu} \nu}_{\phantom{be} \nu} +
		\zeta^{ae \mu \nu} \zeta^{bd}_{\phantom{ae} \nu \mu}
	\right) , 
\label{app_B_180} \\
	\lambda_2 & =g^2 
	f^{amn} f^{npb} \left( 
			\zeta^{ab \phantom{\mu} \nu}_{\phantom{ab} \mu} \zeta^{mp \phantom{\nu} \mu}_{\phantom{mp} \nu} 
			- \zeta^{ab \phantom{\mu} \mu}_{\phantom{ab} \mu} \zeta^{mp \phantom{\nu} \nu}_{\phantom{mp} \nu}
	\right) , 
\label{app_B_190} \\
	\phi_0 & = - g^2 f^{abc} f^{cde} \left( 
		\zeta^{ad \phantom{\mu} \mu}_{\phantom{ad} \mu} \lambda^{be \phantom{\nu} \nu}_{\phantom{be} \nu} +
		\zeta^{ae \mu \nu} \lambda^{bd}_{\phantom{ae} \mu \nu}
	\right) . 
\label{app_B_200}
\end{align}
In Eq.~\eqref{app_B_175}, the symmetry of the quantities $\xi_{\mu \nu}$ follows from the commutativity of the partial derivatives $\partial_\mu \partial_\nu$.

The left-hand side of the contraction \eqref{app_B_20} has the following form: 
\begin{equation}
\begin{split}
	& \left\langle 
		A^m_\mu(y) \left\lbrace \partial_\nu \left[ 
			\partial^\mu A^{m \nu} - \partial^\nu A^{m \mu} 
			+ g f^{m n a} \left( A^{n \mu} A^{a \nu} - A^{a \mu} A^{n \nu} \right) 
	\right]	+ g f^{mna} A^n_\nu F^{a \mu \nu} + g f^{man} A^a_\nu F^{n \mu \nu} 
	\right\rbrace \right\rangle  
\\
	& =\left\langle 
		A^m_\mu(y) \left\lbrace \partial_\nu \left[ 
			\partial^\mu A^{m \nu} - \partial^\nu A^{m \mu} 
			+ g f^{m n a} \left( A^{n \mu} A^{a \nu} - A^{a \mu} A^{n \nu} \right) 
	\right]	
	\right. 
	\right. 
\\
	& \left. \left. 
	+ g f^{mna} A^n_\nu \left( \mathcal{F}^{a \mu \nu} + \mathfrak{F}^{a \mu \nu}\right) 
	+ g f^{man} A^a_\nu \left[ 
		\partial^\mu A^{n \nu} - \partial^\nu A^{n \mu} 
		+ g f^{n p b} \left( A^{p \mu} A^{b \nu} - A^{b \mu} A^{p \nu}\right) 
	\right] \right\rbrace 
	\right\rangle .
\label{app_B_210}
\end{split}
\end{equation}
Following the approximation \eqref{app_B_30}, the terms containing second derivatives can be written in the form:
\begin{equation}
	\expval{A^m_\mu \partial_\nu \left( 
			\partial^\mu A^{m \nu} - \partial^\nu A^{m \mu} 
	\right)} \approx \phi(y) \left[ 
		\zeta^{mm}_{\phantom{mm} \mu \nu} \partial^\mu \partial^\nu \chi - 
		\zeta^{mm \phantom{\mu} \mu}_{\phantom{mm} \mu}\Box \chi 
	\right] . 
\label{app_B_220}
\end{equation}
According to our approximation, all expectation values of products of three field operators vanish:
\begin{equation}
	\expval{A^m_\mu(y) \left( A^{n \mu} A^{a \nu} - A^{a \mu} A^{n \nu}\right)} \approx 0 , 
	\expval{A^m_\mu(y) A^n_\nu \left( 
		\partial^\mu A^{a \nu} - \partial^\nu A^{a \mu}
	\right) } \approx 0, 
	\expval{A^m_\mu(y) A^a_\nu \left( \partial^\mu A^{n \nu} - \partial^\nu A^{n \mu} \right)} 
	\approx 0 . 
\label{app_B_230}
\end{equation}
In our approximation, the expectation values of products of four field operators take the following form:
\begin{align}
	& g^2 f^{mna} f^{abc} \expval{A^m_\mu(y) A^n_\nu A^{b \mu} A^{c \nu}} \approx 
	g^2 f^{mna} f^{abc} \zeta^{mn}_{\phantom{mn} \mu \nu} \zeta^{bc \mu \nu} 
	\chi(y) \chi \phi^2 , 
\label{app_B_240}\\
	& g^2 f^{mna} f^{apq} \expval{A^m_\mu(y) A^n_\nu A^{p \mu} A^{q \nu}} \approx 
	g^2 f^{mna} f^{apq} \left( 
		\zeta^{mn}_{\phantom{mn} \mu \nu} \zeta^{pq \mu \nu} + 
		\zeta^{mp \phantom{\mu} \mu}_{\phantom{mp} \mu} \zeta^{nq \phantom{\nu} \nu}_{\phantom{nq} \nu} + 
		\zeta^{mq \mu \nu} \zeta^{np}_{\phantom{mq} \nu \mu}
	\right) 
	\chi(y) \chi^3  
\nonumber \\ 
& \quad  \, + g^2 f^{mna} f^{apq} \left( 
		\zeta^{mn}_{\phantom{mn} \mu \nu} \lambda^{pq \mu \nu} + 
		\zeta^{mp \phantom{\mu} \mu}_{\phantom{mp} \mu} \lambda^{nq \phantom{\nu} \nu}_{\phantom{nq} \nu} + 
		\zeta^{mq \mu \nu} \lambda^{np}_{\phantom{mq} \nu \mu}
	\right) \chi(y) \chi , 
\label{app_B_250}\\
	& g^2 f^{man} f^{npb} \expval{A^m_\mu(y) A^a_\nu 
	\left( A^{p \mu} A^{b \nu} - A^{b \mu} A^{p \nu}\right) } \approx 
	g^2 f^{man} f^{npb} \left( 
		\zeta^{mp \phantom{\mu} \mu}_{\phantom{mp} \mu} \zeta^{ab \phantom{\nu} \nu} - 
		\zeta^{mp \mu \nu} \zeta^{ab}_{\phantom{ab} \nu \mu} 
	\right) \chi(y) \chi \phi^2 . 
\label{app_B_260}
\end{align}
Finally, we obtain the following expression for the contraction \eqref{app_B_20}:
\begin{equation}
	\chi(y) \left[ 
		\xi_{\mu \nu} \partial^\mu \partial^\nu \chi - \xi \Box \chi 
	+ \left( 
		\xi_1 \chi^2 + \xi_2 \phi^2 - \chi_0
	\right) \chi 
	\right] = -  \left\langle 
		\bar{\psi}\left( x \right) A_\mu^m \left( y \right) \lambda^m \gamma^\mu \psi\left( x \right) 
	\right\rangle . 
\label{app_B_270}
\end{equation}
The following notation has been introduced here: 
\begin{align}
	\xi_{\mu \nu} & = \frac{c^{mm}_{\phantom{mm} \nu \mu} + c^{mm}_{\phantom{mm} \mu \nu}}{2}, 
	\quad \xi = c^{mm \phantom{\mu} \mu}_{\phantom{mm} \mu}, 
\label{app_B_280}\\
	\xi_1 & = g^2 f^{mna} f^{apq} \left( 
		\zeta^{mn}_{\phantom{mn} \mu \nu} \zeta^{pq \mu \nu} + 
		\zeta^{mp \phantom{\mu} \mu}_{\phantom{mp} \mu} \zeta^{nq \phantom{\nu} \nu}_{\phantom{nq} \nu} + 
		\zeta^{mq \mu \nu} c^{np}_{\phantom{mq} \nu \mu}
	\right)  , 
\label{app_B_290}\\
	\xi_2 & = g^2 f^{mna} f^{abc} c^{mn}_{\phantom{mn} \mu \nu} c^{bc \mu \nu} + 
	\left( 
		\zeta^{mp \phantom{\mu} \mu}_{\phantom{mp} \mu} \zeta^{ab \phantom{\nu} \nu} - 
		\zeta^{mp \mu \nu} \zeta^{ab}_{\phantom{ab} \nu \mu} 
	\right) ,
\label{app_B_300}\\
	\chi_0 & = - g^2 f^{man} f^{npb} \left( 
		\zeta^{mp \phantom{\mu} \mu}_{\phantom{mp} \mu} \lambda^{nq \phantom{\nu} \nu}_{\phantom{nq} \nu} + 
		\zeta^{mq \mu \nu} \lambda^{np}_{\phantom{mq} \nu \mu}
	\right) .
\end{align}

\section{$\expval{\hat{A}^a_\mu} \neq 0$, $\expval{\hat{A}^m_\nu} \approx 0$}
\label{nonzero_zero}

In this section, we derive approximate equations for the case in which
$\expval{\hat{A}^a_\mu} \neq 0$, $\hat{A}^a_\mu \in G \subset SU(3)$, and 
$\expval{\hat{A}^m_\nu} \approx 0$, $\hat{A}^m_\nu \in SU(3) / G$. The original equations obtained from the infinite system of Dyson--Schwinger equations \eqref{NP_10}--\eqref{NP_40} take the following form:
\begin{align}
	\expval{\hat F^{a \mu \nu}_{; \nu}(x)} 
	& = - \expval{\hat j^{a \mu}(x)} , 
\label{c_10}\\
	\expval{\hat{A}^m_\mu(y) D_\nu \hat{F}^{m \mu \nu}(x)} & = 
	- \expval{\hat{A}^m_\mu(y) \hat{j}^{m \mu}(x)} , 
\label{c_20}
\end{align}
where 
$	\hat{F}^a_{\mu \nu} = 
	\mathcal{\hat{F}}^a_{\mu \nu} + g f^{a m n} 
		\hat{A}^m_\mu \hat{A}^n_\nu 
	=  \mathcal{\hat{F}}^a_{\mu \nu} + \mathfrak{\hat{F}}^a_{\mu \nu} 
$ and the derivatives act on functions of the coordinate $x$, and henceforth we suppress the dependence on the coordinate $x$, retaining the dependence of the functions and operators on the coordinate $y$. For our approximation, $\expval{\hat{A}^a_\mu} \approx A^a_\mu$, and to approximately describe the two-point Green's function, we introduce the scalar field $\chi$ as follows:
\begin{align}
	\expval{\hat{A}^m_\mu \hat{A}^n_\nu} & \approx \zeta^{mn}_{\phantom{mn} \mu \nu} \chi^2 , 
\label{c_25} \\
	\expval{\hat{A}^m_\mu(y) \hat{A}^n_\nu} & \approx 
	\zeta^{mn}_{\phantom{mn} \mu \nu} \chi(y) \chi , 
\label{c_30}
\end{align}
where $\zeta^{mn}_{\phantom{mn} \mu \nu}$ are some constants. 

It is very important to note the following fact. Since we consider the case $\expval{\hat{A}^a_\mu} \neq 0$, this implies that the physical systems described in this way possess preferred directions both in color space and in spacetime. For example, this occurs when a longitudinal color-electric field emerges in a flux tube stretched between a quark and an antiquark, or when a $Y$-shaped color-electric field forms between three quarks, as occurs in nucleons. This means that the coefficients $\zeta^{mn}_{\phantom{mn} \mu \nu}$ are not symmetric with respect to either the color or spacetime indices.

To evaluate the left-hand side of the contraction \eqref{c_20}, we need an expression for the quantity
$
\expval{\hat{A}^m_\mu \partial_\rho \hat{A}^n_\nu}.
$
To obtain this expression, we consider \eqref{c_30} with the corresponding limiting transition:
\begin{equation}
	\expval{\hat{A}^m_\mu(y) \partial_\rho \hat{A}^n_\nu(x)}_{y \rightarrow x} \approx 
	\zeta^{mn}_{\phantom{mn} \mu \nu} \chi(x) \partial_\rho \chi(x) . 
\label{c_35}
\end{equation}

The left-hand side of Eq. \eqref{c_10} looks like
\begin{equation}
\begin{split}
	\expval{\hat F^{a \mu \nu}_{; \nu}} & = 
	\widetilde{D}_\nu	\mathcal{F}^{a \mu \nu} 
	+ \widetilde{D}_\nu \mathfrak{F}^{a \mu \nu} 
	+ g f^{amn} \expval{\hat{A}^m_\nu \hat{F}^{n \mu \nu}} = 
	\widetilde{D}_\nu	\mathcal{F}^{a \mu \nu} 
	+ g f^{amn} \widetilde{D}_\nu \left( \zeta^{mn \mu \nu} \chi^2 \right) 
\\
	& + g f^{amn} \expval{ \hat{A}^m_\nu 
	\left[ 
		\partial^\mu \hat{A}^{n \nu} - \partial^\nu \hat{A}^{n \mu} 
		+ g f^{npb} \left( \hat{A}^{p \mu} A^{b \nu} - A^{b \mu} \hat{A}^{p \nu} \right) 
		\right] 
	}
\\
	& \approx \widetilde{D}_\nu	\mathcal{F}^{a \mu \nu} 
	+ g f^{amn} \widetilde{D}_\nu \left( \zeta^{mn \mu \nu} \chi^2 \right) 
	+ g f^{amn} \left( 
		\zeta^{mn \phantom{\nu} \nu}_{\phantom{mn} \nu} \partial^\mu \chi - 
		\zeta^{mn \phantom{\nu} \mu}_{\phantom{mn} \nu} \partial^\nu \chi 
	\right) \chi  
\\
	& + g^2 f^{amn} f^{npb} \left( 
		\zeta^{mp \phantom{\nu} \mu}_{\phantom{mp} \nu} A^{b \nu} 
		- \zeta^{mp \phantom{\nu} \nu}_{\phantom{mp} \nu} A^{b \mu}
	\right) \chi^2 
\\
	& = \widetilde{D}_\nu	\mathcal{F}^{a \mu \nu} 
	+ g \widetilde{D}_\nu \left( \zeta^{a \mu \nu} \chi^2 \right) 
	- \frac{g}{2} \widetilde{D}_\nu \left( \zeta^{a \nu \mu} \chi^2 \right) 
	- \frac{g}{2} \widetilde{D}^\mu \left( \zeta^{a} \chi^2 \right) 
\\
	& + g^2 \left[ 
		\frac{1}{2} f^{abc} A^b_\nu \zeta^{c \nu \mu} 
		- \frac{1}{2} f^{abc} A^{b \mu} \zeta^{c} 
	 + f^{amn} f^{npb} \left( 
			\zeta^{mp \phantom{\nu} \mu}_{\phantom{mp} \nu} A^{b \nu} 
			- \zeta^{mp \phantom{\nu} \nu}_{\phantom{mp} \nu} A^{b \mu}
	\right)
	\right] \chi^2 .
\end{split}
\label{c_40}
\end{equation}
Here 
$
	\widetilde{D}_\nu	(\cdots) = \partial_\nu (\cdots) + g f^{abc} A^b_\nu (\cdots)
$ is a gauge derivative in the subgroup $G \subset SU(3)$; 
$
	\expval{\mathcal{\hat{F}}^a_{\mu \nu}} \approx \mathcal{F}^{a}_{\mu \nu} 
$, since in our approximation
$
	\expval{\hat{A}^a_\mu} \approx \hat{A}^a_\mu
$; $
	\zeta^a = f^{amn} \zeta^{mn \phantom{\nu} \nu}_{\phantom{mn} \nu}
$; 
$
	\zeta^{a \mu \nu} = f^{amn} \zeta^{mn \mu \nu}
$. 

Collecting in \eqref{c_40} the terms containing $\chi^2$ and $\partial_\mu \chi$, we can write Eq.~\eqref{c_10} in the following form:
\begin{equation}
	\widetilde{D}_\nu	\mathcal{F}^{a \mu \nu} 
	+ g \widetilde{D}_\nu \left( \zeta^{a \mu \nu} \chi^2 \right) 
	- \frac{g}{2} \widetilde{D}_\nu \left( \zeta^{a \nu \mu} \chi^2 \right) 
	- \frac{g}{2} \widetilde{D}^\mu \left( \zeta^{a} \chi^2 \right) 
	+ \xi^{a b \mu \nu} A^b_\nu \chi^2 
	= - \left\langle \hat{\bar{\psi}} \gamma^\mu \lambda^a \hat{\psi} \right\rangle , 
\label{c_60}
\end{equation}
where 
$
	\xi^{ab \mu}_{\phantom{ab} \nu} = g^2 \left[ 
		\frac{1}{2} f^{abc} \zeta^{c \nu \mu} 
		- \frac{1}{2} f^{abc} \eta^{\mu \nu} \zeta^{c} 
		+ f^{amn} f^{npb} \left( 
		\zeta^{mp \nu \mu}
		- \zeta^{mp \phantom{\alpha} \alpha}_{\phantom{mp} \alpha} \eta^{\mu \nu} 
	\right) 
	\right] 
$ are the coefficients expressed in terms of the structure constants $f^{ABC}$ and the approximation parameters $\zeta^{mn \mu \nu}$. The presence of the term $A^b_\nu$ in \eqref{c_60} indicates that, within this approximation, the ``almost classical'' degrees of freedom are described by a Proca-like theory interacting with the scalar field $\chi$.

To derive the equation for the scalar function $\chi$, let us consider the contraction 
\begin{equation}
\begin{split}
	& \expval{\hat{A}^m_\mu(y) D_\nu \hat{F}^{m \mu \nu}(x)} = 
	\left\langle 
		\hat{A}^m_\mu(y) \left\lbrace \partial_\nu \left[ 
			\partial^\mu \hat{A}^{m \nu} - \partial^\nu \hat{A}^{m \mu} 
			+ g f^{m n a} \left( \hat{A}^{n \mu} A^{a \nu} - A^{a \mu} \hat{A}^{n \nu} \right) 
	\right]	
	\right. \right. 
\\
	& \left. \left. 
	+ g f^{mna} \hat{A}^m_\mu(y) \hat{A}^n_\nu \left( 
		\mathcal{\hat{F}}^{a \mu \nu} + \mathfrak{F}^{a \mu \nu} 
	\right)  
	+ g f^{man} \hat{A}^m_\mu(y) A^a_\nu \left[ 
		\partial^\mu \hat{A}^{n \nu} - \partial^\nu \hat{A}^{n \mu} 
		+ g f^{n p b} \left( \hat{A}^{p \mu} A^{b \nu} - A^{b \mu} \hat{A}^{p \nu} \right) 
	\right] 
	\right\rbrace 
	\right\rangle 
\\
	& \approx \zeta^{mm \mu \nu} \chi(y)\partial_\mu \partial_\nu \chi 
	- \zeta^{mm \phantom{\mu} \mu}_{\phantom{mm} \mu} \chi(y) \Box \chi 
	+ g f^{mna} \zeta^{mn \phantom{\mu} \mu}_{\phantom{mn} \mu} 
	\chi(y) \partial_\nu \left( A^{a \nu} \chi\right) 
	- g f^{mna} \zeta^{mn \mu \nu} \chi(y) \partial_\nu \left( A^a_\mu \chi\right) 
\\
	& + g f^{mna} \zeta^{mn}_{\phantom{mn} \mu \nu} \mathcal{F}^{a \mu \nu} \chi(y) \chi 
	+ g^2 f^{mna} f^{apq} \expval{\hat{A}^m_\mu(y) \hat{A}^n_\nu \hat{A}^{p \mu} \hat{A}^{q \nu}} 
\\
	& - g f^{amn} \zeta^{mn \mu \nu} A^a_\nu \chi(y) \partial_\mu \chi 
	+ g f^{amn} \zeta^{mn \phantom{\mu} \mu}_{\phantom{mn} \mu} A^{a \nu} \chi(y) \partial_\nu \chi 
 + g^2 f^{man} f^{npb} \left( 
		\zeta^{mp \phantom{\mu} \mu}_{\phantom{mp} \mu} A^a_\nu A^{b \nu} 
		- \zeta^{mp \mu \nu} A^a_\nu A^b_\mu 
	\right) \chi(y) \chi . 
\label{c_70}
\end{split}
\end{equation}
The term
$f^{mna} f^{apq} \expval{\hat{A}^m_\mu(y) \hat{A}^n_\nu \hat{A}^{p \mu} \hat{A}^{q \nu}}$
is evaluated using \eqref{app_B_90}:
\begin{equation}
\begin{split}
	f^{mna} f^{apq} \expval{\hat{A}^m_\mu(y) \hat{A}^n_\nu \hat{A}^{p \mu} \hat{A}^{q \nu}} 
	\approx & f^{mna} f^{apq} \chi(y) \left[ 
	\left( 
		\zeta^{mn}_{\phantom{mn} \mu \nu} \zeta^{pq \mu \nu} + 
		\zeta^{mp \phantom{\mu} \mu}_{\phantom{mp} \mu} \zeta^{nq \phantom{\nu} \nu}_{\phantom{mp} \nu} + 
		\zeta^{mq \phantom{\nu} \nu}_{\phantom{mq} \mu} \zeta^{np \phantom{\nu} \mu}_{\phantom{np} \nu} 
	\right) \chi^3
	\right. 
\\
	& \left. \left( 
		\zeta^{mn}_{\phantom{mn} \mu \nu} \lambda^{pq \mu \nu} + 
		\zeta^{mp \phantom{\mu} \mu}_{\phantom{mp} \mu} 
		\lambda^{nq \phantom{\nu} \nu}_{\phantom{mp} \nu} + 
		\zeta^{mq \phantom{\nu} \nu}_{\phantom{mq} \mu} 
		\lambda^{np \phantom{\nu} \mu}_{\phantom{np} \nu} 
	\right) \chi 
	\right]  .
\end{split}
\label{c_72}
\end{equation}
All of this leads to the following expression: 
\begin{equation}
\begin{split}
	\expval{\hat{A}^m_\mu(y) D_\nu \hat{F}^{m \mu \nu}(x)} 
	& \approx \chi(y) \left\lbrace 
		\zeta^{\mu \nu} \partial_\mu \partial_\nu \chi 
		+ g \zeta^{a \mu \nu} \mathcal{F}^a_{\mu \nu} \chi 
		+ g \nabla_\nu 
		\left( 
			 \zeta^a A^{a \nu} \chi - \zeta^{a \mu \nu} A^a_\mu \chi  
		\right) 
		\right. 
\\
		& \left. 
		- g 
		\left( 
		\zeta^{a \mu \nu} A^a_\nu - \zeta^a A^{a \mu} 
		\right) \nabla_\mu \chi 
		+ \Lambda \left( 
			\chi^2 - \chi_0
		\right) \chi 
		- m^{ab \mu \nu} A^a_\nu A^b_\mu \chi 
	\right\rbrace . 
\end{split}
\label{c_75}
\end{equation}
Here 
$
	\zeta^{\mu \nu} = 
	\zeta^{m m \mu \nu} - \zeta^{mm \phantom{\alpha} \alpha}_{\phantom{mm} \alpha} \eta^{\mu \nu}
$ and the following coefficients have been introduced:
\begin{align}
	\Lambda & = g^2 f^{mna} f^{apq} \left( 
		\zeta^{mn}_{\phantom{mn} \mu \nu} \zeta^{pq \mu \nu} + 
		\zeta^{mp \phantom{\mu} \mu}_{\phantom{mp} \mu} \zeta^{nq \phantom{\nu} \nu}_{\phantom{mp} \nu} + 
		\zeta^{mp \phantom{\mu} \mu}_{\phantom{mp} \mu} 
		\zeta^{nq \phantom{\nu} \nu}_{\phantom{mp} \nu} + 
		\zeta^{mq \phantom{\nu} \nu}_{\phantom{mq} \mu} 
		\zeta^{np \phantom{\nu} \mu}_{\phantom{np} \nu}  
	\right) ,
\nonumber \\
	\chi_0 & = - \frac{g^2}{\Lambda} f^{mna} f^{apq} \left( 
		\zeta^{mn}_{\phantom{mn} \mu \nu} \lambda^{pq \mu \nu} + 
		\zeta^{mp \phantom{\mu} \mu}_{\phantom{mp} \mu} 
		\lambda^{nq \phantom{\nu} \nu}_{\phantom{mp} \nu} + 
		\zeta^{mq \phantom{\nu} \nu}_{\phantom{mq} \mu} 
		\lambda^{np \phantom{\nu} \mu}_{\phantom{np} \nu} 
	\right) , 
\nonumber \\
	m^{ab \mu \nu} & = g^2 f^{amn} f^{bpn} \left( 
		\zeta^{mp \phantom{\alpha} \alpha}_{\phantom{mp} \alpha} \eta^{\mu \nu} 
		- \zeta^{mp \mu \nu}
	\right) .
\nonumber 
\end{align}
The quantities $m^{ab \mu \nu}$ may be referred to as the mass matrix.

Ultimately, we obtain the contraction that determines the scalar field $\chi$, which approximately describes the two-point Green's function \eqref{c_25}:
\begin{equation}
\begin{split}
& \chi(y) \left[ 
		\zeta^{\mu \nu} \partial_\mu \partial_\nu \chi 
		+ g \zeta^{a \mu \nu} \mathcal{F}^a_{\mu \nu} \chi 
		+ g \nabla_\nu 
		\left( 
			 \zeta^a A^{a \nu} \chi - \zeta^{a \mu \nu} A^a_\mu \chi  
		\right) 
		\right. 
\\
		& \left. 
		- g 
		\left( 
		\zeta^{a \mu \nu} A^a_\nu - \zeta^a A^{a \mu} 
		\right) \nabla_\mu \chi 
		+ \Lambda \left( 
			\chi^2 - \chi_0
		\right) \chi 
		- m^{ab \mu \nu} A^a_\nu A^b_\mu \chi 
	\right]  = 
	- \expval{\hat{A}^m_\mu(y) \hat{\bar{\psi}} \gamma^\mu \lambda^m \hat{\psi}} . 
\end{split}
\label{c_80}
\end{equation}

In conclusion to this section, we note that the quantities $\zeta^{mn}_{\phantom{mn} \mu \nu}$ can be substantially simplified by factorizing them in the following form:
$\zeta^{mn}_{\phantom{mn} \mu \nu} = l^m_\mu l^n_\nu$. That is
\begin{equation}
	\expval{\hat{A}^m_\mu(y) \hat{A}^n_\nu} \approx 
	l^m_\mu l^n_\nu \chi(y) \chi . 
\label{c_90}
\end{equation}
Physically, this means that the two-point Green's function is determined by a certain color vector $l^m_\mu$, which characterizes the color and spacetime asymmetry induced by the color fields between the quarks, and by the scalar field $\chi$. This approximation can be understood as follows. The operator $\hat{A}^m_\mu$ is a complicated function of the position in spacetime; nevertheless, the behavior of the two-point Green's function can be approximately represented as a product of a vector $l^m_\mu$, carrying the same indices $m,\mu$, and a scalar field $\chi$, which approximately describes the behavior of all components of $\hat{A}^m_\mu$.

\section{Approximation for the Dirac equation and currents}
\label{Dirac_eqn}

In this section, we derive an approximation for the Dirac equation
\begin{equation}
	\expval{
	\hat{\bar{\psi}}_{\gamma k}(y) 
	\left[ 
		\imath \qty(\gamma^\mu)_{\alpha \beta}
		\qty(D_\mu)_{ij} \hat{\psi}_{\beta j} - m \hat{\psi}_{\alpha i}
	\right] 
	} = 0 ,
\label{В_10}
\end{equation}
where the fermion field $\hat{\psi}$ describes quarks, 
$
	D_\mu \psi_{\alpha i}= 
	\partial_\mu \psi_{\alpha i} - \imath \frac{g}{2} \qty(\lambda^B)_{ij} 
	\qty(\gamma^\mu)_{\alpha \beta} A^B_\mu \psi_{\beta j} 
$ is the covariant derivative of the spinor; $\alpha, \beta$ are spinor indices, while $i, j$ are indices of the spinor triplet $\psi$. For clarity, we restore the operator notation. The operators entering Eq.~\eqref{В_10} can be written in the following form: 
\begin{equation}
	\hat{\psi}_{\alpha i} = \vartheta_{\alpha i} + \widehat{\delta \psi}_{\alpha i}, \quad 
	\hat{A}^a_\mu \approx A^a_\mu , \quad 
	\expval{\hat{A}^m_\mu} \approx 0, 
\label{В_20}
\end{equation}
where $\vartheta = \expval{\psi}$ is the expectation value of the fermion field, $\widehat{\delta \psi}$ are quantum fluctuations (sea quarks). 

We can now specify all the approximations that will be used to derive the Dirac equation within our approximation scheme:
\begin{align}
	\expval{\widehat{\delta \bar{\psi}}_{\gamma k}(y) \widehat{\delta \psi}_{\alpha i}} & \approx 
	\bar{\varsigma}_{\gamma k}(y) \varsigma_{\alpha i}, 
\label{В_30}\\
	\expval{\hat{\bar{\psi}}_{\gamma k}(y) \hat{\psi}_{\alpha i}} & \approx 
	\bar{\vartheta}_{\gamma k}(y) \vartheta_{\alpha i}
	+ \bar{\varsigma}_{\gamma k}(y) \varsigma_{\alpha i}, 
\label{В_40}\\
	\expval{
		A^m_\mu \widehat{\delta \psi}_{\alpha i}} 
		= \expval{\widehat{\delta \bar{\psi}} A^m_\mu}
		& \approx 0 , 
\label{В_60}\\
	\expval{
		\widehat{\delta \bar{\psi}}_{\gamma k}(y) 
		A^a_\mu \qty( \gamma^\mu)_{\alpha \beta} \lambda^a_{ij} 
		\widehat{\delta \psi}_{\beta j}
	} 
	& = A^a_\mu \expval{
		\widehat{\delta \bar{\psi}}_{\gamma k}(y) 
		\qty( \gamma^\mu)_{\alpha \beta} \lambda^a_{ij} 
		\widehat{\delta \psi}_{\beta j}
	} \approx \bar{\varsigma}_{\gamma k}(y) 
	\qty( \gamma^\mu)_{\alpha \beta} A^a_\mu \lambda^a_{ij} \varsigma_{\beta j} , 
\label{В_65}\\
	\expval{
		\widehat{\delta \bar{\psi}}_{\gamma k}(y) 
		A^m_\mu\qty( \gamma^\mu)_{\alpha \beta} \lambda^m_{ij} 
		\widehat{\delta \psi}_{\beta j}
	} 
		& \approx 
		\kappa_1 \bar{\varsigma}_{\gamma k}(y) \chi \varsigma_{\beta j} 
		+ \kappa_2 \bar{\varsigma}_{\gamma k}(y) \varsigma_{\alpha i}  
		\bar{\varsigma}_{\beta j} \varsigma_{\beta j} , 
\label{В_70}\\
	\expval{
		\hat{\bar{\psi}}_{\gamma k}(y) A^B_\mu\qty( \gamma^\mu)_{\alpha \beta} 
		\lambda^B_{ij} \hat{\psi}_{\beta j} 
	} & = 
	\expval{ \left(\bar{\vartheta}_{\gamma k}(y) + \widehat{\delta \bar{\psi}}_{\gamma k}(y) \right) 
		A^B_\mu\qty( \gamma^\mu)_{\alpha \beta} \lambda^B_{ij} 
		\left( \vartheta_{\beta j} + \widehat{\delta \psi}_{\beta j} \right) 
	}
\nonumber \\
	\approx \bar{\vartheta}_{\gamma k}(y) A^a_\mu\qty( \gamma^\mu)_{\alpha \beta} 
	\lambda^a_{ij}  \vartheta_{\beta j} 
	& + \widehat{\delta \bar{\psi}}_{\gamma k}(y) A^a_\mu\qty( \gamma^\mu)_{\alpha \beta} 
		\lambda^a_{ij}  \widehat{\delta \psi}_{\beta j} 
	+ \widehat{\delta \bar{\psi}}_{\gamma k}(y) A^m_\mu\qty( \gamma^\mu)_{\alpha \beta} 
		\lambda^m_{ij}  \widehat{\delta \psi}_{\beta j} 
\nonumber \\
	\approx \bar{\vartheta}_{\gamma k}(y) A^a_\mu\qty( \gamma^\mu)_{\alpha \beta} 
		\lambda^a_{ij}  \vartheta_{\beta j} 
	& + \bar{\varsigma}_{\gamma k}(y) A^a_\mu\qty( \gamma^\mu)_{\alpha \beta} 
	\lambda^a_{ij}  \varsigma_{\beta j} 
	+ \kappa_1 \bar{\varsigma}_{\gamma k}(y) \chi \varsigma_{\alpha i} + 
	\kappa_2 \left( \bar{\varsigma}_{\gamma k}(y) \varsigma_{\alpha i} \right) 
	\left( \bar{\varsigma}_{\beta j} \varsigma_{\beta j} \right) . 
\label{В_80}
\end{align}
In the approximate expression for the right-hand side of \eqref{В_70}, only nonzero expectation values of certain combinations of the operators $A^B_\mu$ and $\psi$ should be retained:
$	\expval{A^a_\mu A^{a \mu}} $ and 
$
	\left[ \bar{\varsigma}(y) \varsigma\right]\left( \bar{\varsigma} \varsigma\right)
$. 
This expression is essentially the simplest Lorentz-invariant expression containing quadratic combinations of the gauge field and the spinor $\psi$ describing the quarks. We have also used the obvious identity
$
	\expval{\varsigma_{\gamma k}(y) A^m_\mu \varsigma_{\beta j}} = 0
$ and assumed that
$
	\expval{\widehat{\delta \psi}_{\gamma k}(y) A^m_\mu} \approx 0 
$ and 
$
	\expval{A^m_\mu \widehat{\delta \psi}_{\beta j}} \approx 0 
$. 

Substituting the expansion for $\hat \psi$ from \eqref{В_20} into the Dirac equation \eqref{В_10} and using the approximations \eqref{В_30}--\eqref{В_80}, we obtain an equation containing two unknown and mutually independent functions $\psi$ and $\varsigma$. For this equation to be satisfied, the expressions containing only $\vartheta$ and $\varsigma$ must be set equal to zero separately, which leads to two equations:
\begin{align}
	\bar{\vartheta}_{\gamma k}(y) 
	\left[ 
		\imath \qty(\gamma^\mu)_{\alpha \beta}
		\qty(\tilde{D}_\mu)_{ij} \vartheta_{\beta j} - m \vartheta_{\alpha i}
	\right] & = 0 ,
\label{В_90}\\
	\bar{\varsigma}_{\gamma k}(y) \left[ 
		\imath \qty(\gamma^\mu)_{\alpha \beta}
			\qty(\tilde{D}_\mu)_{ij} \varsigma_{\beta j} - 
			\left( 
				m - \kappa_1 \chi 
			\right) \varsigma_{\alpha i} + \kappa_2 \left( \bar{\varsigma} \varsigma \right) \varsigma_{\alpha i} 
	\right] & = 0 . 
\label{В_100}
\end{align}
In deriving these equations, we have taken into account that the covariant derivative $D_\mu$ contains the terms $\lambda^a A^a_\mu$ and $\lambda^m A^m_\mu$, as well as the approximations \eqref{В_30}--\eqref{В_80}.

Finally, after canceling the common factors $\bar{\vartheta}{\gamma k}(y)$ and $\bar{\varsigma}{\gamma k}(y)$, we obtain the following pair of equations: 
\begin{align}
	\imath \qty(\gamma^\mu)_{\alpha \beta}
	\qty(\tilde{D}_\mu)_{ij} \vartheta_{\beta j} - m \vartheta_{\alpha i} & = 0 ,
\label{В_110}\\
	\imath \qty(\gamma^\mu)_{\alpha \beta}
	\qty(\tilde{D}_\mu)_{ij} \varsigma_{\beta j} - 
	\left( 
		m - \kappa_1 \chi 
	\right) \varsigma_{\alpha i} + \kappa_2 \left( \bar{\varsigma} \varsigma \right) \varsigma_{\alpha i}  
	& = 0 . 
\label{В_120}
\end{align}
To obtain approximations for the currents $\expval{\hat{\bar{\psi}} \gamma^\mu \lambda^a \hat{\psi}}$ and $\expval{\hat{A}^m_\mu(y) \hat{\bar{\psi}} \gamma^\mu \lambda^m \hat{\psi}}$, we use the following approximations:
\begin{align}
 \expval{
 	\hat{\bar{\psi}}_{\alpha i} \qty( \gamma^\mu)_{\alpha \beta} \lambda^a_{ij}  \hat{\psi}_{\beta j}
 } & = 
	\expval{
		\left(\bar{\vartheta}_{\alpha i} + \widehat{\delta \bar{\psi}}_{\alpha i}(y) \right) 
		\qty( \gamma^\mu)_{\alpha \beta} \lambda^a_{ij} 
		\left( \vartheta_{\beta j} + \widehat{\delta \psi}_{\beta j} \right) 
	}
	\approx 
	\bar{\vartheta}_{\alpha i} \qty( \gamma^\mu)_{\alpha \beta} \lambda^a_{ij}  \vartheta_{\beta j} 
	+ \bar{\varsigma}_{\alpha i} \qty( \gamma^\mu)_{\alpha \beta} \lambda^a_{ij}  \varsigma_{\beta j} , 
\label{В_130}\\
	\expval{
		\hat{\bar{\psi}}_{\alpha i} A^m_\mu(y) \qty( \gamma^\mu)_{\alpha \beta} 
		\lambda^m_{ij} \hat{\psi}_{\beta j} 
	} & = 
	\expval{ \left(\bar{\vartheta}_{\alpha i} + \widehat{\delta \bar{\psi}}_{\alpha i}(y) \right) 
		A^m_\mu(y) \qty( \gamma^\mu)_{\alpha \beta} \lambda^m_{ij} 
		\left( \vartheta_{\beta j} + \widehat{\delta \psi}_{\beta j} \right) 
	} 
	\approx 
	\kappa_1 \chi (y)\bar{\varsigma} \varsigma . 
\label{В_140}
\end{align}
Here 
$
	\bar{\varsigma} \varsigma = \bar{\varsigma}_{\alpha i} \varsigma_{\alpha i}  
$.

\end{document}